\documentclass{aa}  

\usepackage{graphicx}
\usepackage{txfonts}
\usepackage{xcolor}
\usepackage{ulem}
\newcommand*{\spr}{{\sc Spritz}}
\newcommand*{\cii}{{[$\ion{\rm C}{II}$]\,}}
\newcommand*{\hers}{\textit{Herschel\,}}
\defcitealias{Polletta2007}{P07}
\defcitealias{Gruppioni2013}{G13}
\defcitealias{Vallini2015}{V15}
\defcitealias{Bisigello2021}{B21}
\usepackage[flushleft]{threeparttable}
\usepackage{scrextend}
\usepackage{multirow}

\begin{document}

   \title{\spr{} is sparkling:  simulated CO and \cii luminosities}

   \author{L. Bisigello
          \inst{1,2,3}\thanks{laura.bisigello@unipd.it} \and
          L. Vallini\inst{4} \and
          C. Gruppioni\inst{3} \and 
          F. Esposito\inst{5,3} \and
          F. Calura\inst{3} \and
          I. Delvecchio\inst{6} \and
          A. Feltre\inst{3} \and
          F. Pozzi\inst{5,3} \and
          G. Rodighiero\inst{1,2}
          }

   \institute{Dipartimento di Fisica e Astronomia, Università di Padova, Vicolo dell'Osservatorio, 3, I-35122, Padova, Italy,
   \and
   INAF-Osservatorio Astronomico di Padova, vicolo dell'Osservatorio 5, I-35122, Padova, Italy,
   \and
   INAF-Osservatorio di Astrofisica e Scienza dello Spazio, via Gobetti 93/3, I-40129, Bologna, Italy
   \and
   Scuola Normale Superiore, Piazza dei Cavalieri 7, I-56126 Pisa, Italy
   \and
   Dipartimento di Fisica e Astronomia, Università degli Studi di Bologna, Via P. Gobetti 93/2, I-40129 Bologna, Italy \and
   INAF - Osservatorio Astronomico di Brera, Via Brera 28, 20121, Milano, Italy
}

   \date{Received ; accepted }

 
  \abstract
   {}
   {We present a new prediction of the luminosity functions of the \cii line at 158 $\mu$m, of the CO lines from $J=0$ to $J=24$, and of the molecular gas mass density up to $z=10$, using the Spectro-Photometric Realisations of Infrared-selected Targets at all-z (\spr) simulation (Bisigello et al. 2021).}  
   {We update the state-of-the-art phenomenological simulation \spr{} to include both the CO ($J\leq24$) and \cii line luminosities. This has been performed using different empirical and theoretical relations to convert the total infrared luminosity (or star formation rate) to \cii or CO luminosity. The resulting line luminosity functions have been compared for validation with a large set of observations available in the literature. We then used the derived CO and \cii line luminosities to estimate the molecular gas mass density and compare it with available observations.}
   {The CO and \cii luminosity functions presented here are well in agreement with all the available observations. In particular, the best results for \cii are obtained deriving the \cii luminosity directly from the star formation rate, but considering a dependence of this relation on the gas metallicity. For all the CO luminosity functions, the estimates favoured by the data are derived considering different relations, depending on the ionisation mechanism dominating each galaxy, i.e. star formation or active galactic nuclei, and, moreover, deriving the $J\geq4$ CO lines directly from the \cii luminosity. However, further data are necessary to fully discriminate between models. Finally, the best agreement with observations of the molecular gas mass density are derived by converting the \cii luminosity to $\rm H_2$ mass, using a \cii-to-$\rm H_2$ conversion $\sim$130 $\rm M_{\odot}/{\rm L}_{\odot}$. All the line luminosity functions, useful for planning and interpreting future observations, are made publicly available.}
   {}


   \maketitle
%

\section{Introduction}
The molecular phase of the interstellar medium (ISM) is the birthplace of stars, and therefore it plays a central role in galaxy evolution \citep[see e.g. the review by][]{Tacconi2020}. The direct detection of molecular hydrogen ($\rm H_2$) in galaxies is hampered by the fact that this molecule, lacking a permanent dipole moment, possesses no corresponding dipolar rotational transitions. The lowest energy transitions of $\rm H_2$ are the purely rotational quadrupole lines that require high temperatures ($T>500-1000\,\rm K$) to be excited \citep{Bolatto2013}. For this reason carbon monoxide (CO), which is the most abundant molecule after $\rm H_2$ and is easily excited even in cold molecular clouds, is usually used to estimate the molecular gas mass (M$_{\rm H_2}$) via the CO(1-0) emission assuming a CO-to-$\rm H_2$ conversion factor, $\alpha_{CO}$ \citep[e.g.][]{Bolatto2013,Decarli2019,Riechers2019}. \par
CO detections at high-$z$ are almost exclusively reported in rare highly star-forming sub-mm galaxies \citep[e.g.][]{Jarugula2021, Dye2022} and quasars \citep[e.g.][]{Carniani2019, Pensabene2021}, albeit with some exceptions \citep[][]{Dodorico2018, Pavesi2019}. With the Atacama Large Millimetre/submillimetre Array (ALMA), the detection of CO from intermediate redshifts ($z\approx 1-2$) has become feasible also for normal star-forming galaxies \citep[][]{Valentino2020} but at $z>5$ it remains extremely time-consuming \citep[e.g.][]{Vallini2018}. This is primarily due to (i) the overall lower metallicity and dust abundance of early galaxies, resulting in CO being easily dissociated \citep[e.g.][]{Madden2020}, and (ii) the effect of the increased cosmic microwave background (CMB) temperature that represents a stronger background against which CO lines are observed \citep{daCunha2013}.\par
Given the difficulties of observing the CO emission in faint galaxies beyond the Local Universe, several works \citep[e.g.][]{Keating2020}
focused on the exploitation of the global CO line emission signal from unresolved sources through the so-called line intensity mapping (LIM) technique. Models of the expected LIM signal require the derivation of the line luminosity-halo mass relation, which has been often obtained through hydrodynamical simulations and semi-analytical models (SAM) \citep[e.g.][]{Lidz2011,Gong2011,Mashian2015,Li2016,Sun2019,Yang2021}. However, this relation can also be inferred from the observed CO luminosity function \citep[LF,][]{Padmanabhan2018} using an abundance matching technique analogous to that assumed for the stellar mass–halo mass relation \citep[e.g.][]{Berhoozi2010}. Nevertheless, an extensive use of this approach has been hampered, up to now, due to the sparse CO observations available, and the resulting huge uncertainties regarding the evolution of the CO LFs.
\par
CO fails, sometimes, in tracing the whole $\rm H_2$ mass, particularly in low-metallicity galaxies where the reduced dust content results in a deeper penetration of far-ultraviolet photons able to dissociate the CO but not the self-shielded $\rm H_2$. The $\rm H_2$ thus survives outside the CO region \citep[e.g.][]{Gnedin2014} in the so-called CO-dark clouds \citep[e.g.][]{Wolfire2010} and can be instead efficiently traced by another (much brighter) proxy of cold gas, i.e. the \cii line at 158 $\mu$m. The \cii is the major coolant of the cold diffuse medium \citep[][]{Wolfire2003}, and dense Photodissociation Regions \citep[PDRs,][for a recent review]{Hollenbach1999, Wolfire2022} associated with molecular clouds. Most importantly, it is now routinely detected in large samples of galaxies at $z>4-5$, such as those targeted by the ALMA Large Program to INvestigate CII at Early Times \citep[ALPINE;][]{LeFevre2020, Gruppioni2020, Yan2020,Loiacono2021}, and Reionization Era Bright Emission Line Survey \citep[REBELS;][]{Bouwens2021}. The \cii line, being a reliable tracer of the total molecular gas mass \citep{Zanella2018,Madden2020,Vizgan2022}, thus represents a fundamental tool to follow the cosmic evolution of the fuel of star formation and better understand how the gas supply in galaxies moderates the star formation rate (SFR) across the Universe history. \par
The study of the molecular gas mass density is fundamental for understanding which physical processes are driving the change in the star formation rate density (SFRD) occurring at cosmic noon \citep[e.g.][]{Madau2014}. Indeed, it is still a matter of debate whether this is due to a lack of cold gas supply, or a lower efficiency in converting the gas into stars, or to the presence of strong outflows preventing the infall of new cold material. The simplest scenario would expect the SFRD mirroring the cold gas evolution, as gas is being consumed by star formation \citep[e.g.][]{Driver2018}. To further improve our understanding on this topic, it is desirable to complement targeted studies with blind measurements to derive the CO (or \cii) LF at different cosmic epochs. In recent years the ASPECS survey \citep[ALMA Spectroscopic Survey in the Hubble Ultra-Deep Field;][]{Walter2016, Decarli2019, Boogaard2020} was designed exactly for this purpose. \par
At the same time, SAM and empirical models of \cii and CO LFs have started providing predictions for the LF evolution and, most importantly, a framework within which interpreting the upcoming data \citep[][]{Obreschkow2009, Lagos2012, Vallini2016, Popping2019}. However, the majority of these models have issues on reproducing the bright-end of the observed CO LFs at $z>1$ \citep{Decarli2019,Riechers2019}, similarly to what observed for other related quantities, such as the total infrared (IR) LF \citep[e.g.][]{Gruppioni2015,AlcaldePampliega2019} or the dust mass \citep[e.g.][]{Pozzetti2000,Calura2017,Magnelli2020,Pozzi2021}. \par
An alternative approach is based on the exploitation of empirical relations to associate the nebular line emission to dark-matter halos, like recently done by \citet{Chung2020} and \citet{Bethermin2022}. The work by \citet{Chung2020} is based on the halo-galaxy connection presented by \citet[][]{Behroozi2019}, which includes the observed UV LFs as constraints, while\citet{Bethermin2022} adopt the stellar mass functions \citep[see also][]{Bethermin2017}. In this paper, we consider a similar empirical approach by extending the work presented in \citet{Vallini2016} to derive the evolution of the \cii and CO LFs, together with the molecular gas mass density. In particular, our work uses different constraints with respect to \citet{Chung2020} or \citet{Bethermin2022}, as it is based on the state-of-the-art Spectro-Photometric Realisations of Infrared-selected Targets at all-\textit{z} \citep[\spr;][hereafter B21]{Bisigello2021}, which uses in input the observed IR LF \citep{Gruppioni2013} and is not linked to any dark-matter only simulation. \par
The paper is organised as follows. The \spr{} simulation is described in detail in Section \ref{sec:SPRITZ}, while in Section \ref{sec:lines} we list all the relations considered to include CO and \cii in \spr. We compare the CO and \cii LFs with observations available in the literature in Section \ref{sec:comp_obs}. In Section \ref{sec:H2} we focus on the molecular gas mass, describing its derivation in \spr{} and comparing it with observations, and we finally report our conclusions in Section \ref{sec:conclusions}. We consider a $\Lambda$CDM cosmology with $H_0=70\,{\rm km}\,{\rm s}^{-1}{\rm Mpc}^{-1} $, $\Omega_{\rm m}=0.27$ and $\Omega_\Lambda=0.73$ and a Chabrier initial mass function \citep{Chabrier2003}.
\section{The \spr{} simulation}\label{sec:SPRITZ}
The CO and \cii line luminosities reported in this paper are obtained from the \spr{} simulation, described in detail by \citetalias{Bisigello2021}, which includes elliptical galaxies, dwarf irregulars, star-forming galaxies, and active galactic nuclei (AGN). \spr{} is derived starting from a set of galaxy stellar mass functions (GSMF) and luminosity functions (LFs), mainly in the infrared (IR). Then, a spectral energy distribution (SED) template is assigned to each simulated galaxy, allowing us to make predictions for several past, current and future facilities covering different ranges of the electromagnetic spectrum, from the X-ray to the IR. By using the SED templates and a set of empirical and theoretical relations, we derived for each simulated galaxy all the main physical properties, including stellar masses and line luminosities. We will now focus on the part of the simulation relevant for this work. \par
In particular, all the simulated galaxies are extracted from:
\begin{itemize}
    \item the IR LF, as derived from \hers observations by \citet{Gruppioni2013}. These LFs are estimated for different galaxy populations including normal star-forming galaxies (hereafter spiral), starburst (SB) and two composite systems (SF-AGN and SB-AGN). The latter describes two populations with an AGN component that is, however, not the dominant source of power, except in the mid-IR and, partially, in the X-ray. In particular, SF-AGN contains an intrinsically faint AGN hosted by a star-forming galaxy, while SB-AGN host a bright, but heavily obscured AGN hosted by a SB. 
    All these LFs are extrapolated at $z>3$, where \hers observations are not available, by assuming a constant characteristic luminosity (L$^{*}$) and by decreasing the number density at the knee ($\Phi^{*}$) as $\propto(1 +\textit{z})^{k_{\Phi}}$. For the power-law exponent $k_{\Phi}$, we considered a range of values from $-$4 to $-$1 to span different possible scenarios.  
    \item the IR LF of AGN-dominated systems (un-obscured AGN 1 and obscured AGN 2) derived by \citetalias{Bisigello2021} starting from \hers observations complemented by far-ultraviolet observations up to $z=5$ \citep[e.g.][]{Croom2009,McGreer2013,Ross2013,Akiyama2018,Schindler2019}. The LF is described by a modified Schechter function and its evolution has been extrapolated at $z>5$ following the observations at lower-\textit{z}  (i.e. $\Phi^{*}\propto(1 +\textit{z})^{-2.75}$).
    \item the K-band LF of elliptical galaxies (hereafter, Ell), estimated by averaging the LFs by \citet{Arnouts2007}, \citet{Cirasuolo2007} and \citet{Beare2019}. At $z>2$ the LF is extrapolated by keeping the characteristic luminosity constant and decreasing the number density at the knee as $\propto$(1+\textit{z})$^{-1}$. This assumption has little impact, as the number density of elliptical galaxies at $z=2$ is already quite low, and it will be tested with future observations.
    \item the GSMF of dwarf irregular galaxies (hereafter, Irr) derived by \citet{Huertas-Company2016} and complemented by the local GSMF of irregular galaxies observed in the Galaxy And Mass Assembly (GAMA) survey \citep{Moffett2016}. More details are given in Appendix \ref{sec:newspritz}.
\end{itemize}
As mentioned before, we derived for each simulated galaxy the main physical properties (e.g. stellar mass, star formation rate, stellar metallicity, and luminosities). In particular, for the majority of them the total IR luminosity L$_{IR}$ was directly taken from the best template used to derive the observed IR LF. For Irr and Ell, instead, L$_{IR}$ was obtained starting from the galaxy stellar mass or the K-band luminosity assuming a SED template \citep[][for dwarf elliptical and irregulars galaxies, respectively]{Polletta2007,Bianchi2018}. \par
The IR component of the SFR was derived from the L$_{IR}$ assuming the \citet{Kennicutt1998} conversion, while the UV component was derived, assuming the \citet{Kennicutt1998a} relation, from the luminosity at 1600$\,\AA$ not corrected by dust absorption. The 1600$\,\AA$ luminosity is derived directly from the SED template associated to each simulated galaxy, which depends on the galaxy population to which it belongs, and it is taken from a set of 35 empirical templates \citep{Polletta2007,Rieke2009,Gruppioni2010,Bianchi2018}.  These empirical templates are of low-z galaxies, but they represent a good description of galaxies observed by \hers up to $z=3.5$ \citep{Gruppioni2013}. The same procedure  applied to derive the UV luminosity is performed to obtain the galaxy stellar mass, as each template is normalised to 1$\,\rm M_{\odot}$. Finally, the stellar metallicity is  derived from the galaxy stellar mass assuming the mass-metallicity relation by \citet{Wuyts2014}:
\begin{equation}\label{eq:met}
\begin{aligned} 
    &12 + {\rm log}_{10}(O/H) = Z_{0} + {\rm log}_{10} [1-exp(-(M^{*}/M_{0})^{\gamma} ], \\
    &{\rm log}_{10}(M_{0}/M_{\odot}) = (8.86 \pm 0.05) + (2.92 \pm 0.16)\,{\rm log}_{10}(1 + z).
\end{aligned}
\end{equation}
where the asymptotic metallicity is $Z_{0}=$8.69 and the power-law slope at low metallicity is $\gamma=$0.40. 

\par
The \spr{} simulation is in agreement with a large set of observations, ranging from number counts at different wavelengths to the total GSMF \citepalias[see Section 4 in ][]{Bisigello2021}. Of particular interest for this work is the agreement with the observed IR LF at $z\sim5$ \citep{Gruppioni2020}, which was not included as input in the simulation. In particular, among the different high-$z$ extrapolations tested, the best agreement is present when assuming that the number density at the knee of the IR LF evolves as $\propto(1 +\textit{z})^{-1}$ for spiral, SB, SF-AGN and SB-AGN. This agreement supports the validity of the extrapolation performed at $z>3$. \par
On the contrary, some tensions is present between \spr{} and the observed UV LF \citepalias[see Figure 20 in][]{Bisigello2021}. In particular, the bright-end (M$_{1600\text{\AA}}<-22.5$ at $z=0.5$ and -21.0 at $z=1.5$) of the galaxy UV LF is overestimated at $z\leq1.5$, while the faint-end is underestimated, particularly at $z>2$. \par
When looking at the SFR-M$^{*}$ plane in \spr, star-forming galaxies correctly populate the galaxy main sequence, whose normalisation increases with increasing redshift, as also visible in observations \citep[e.g.][]{Noeske2007,Speagle2014,Bisigello2018}. SB are correctly placed above the main sequence, while elliptical galaxies, by construction, are placed below. However, at $z>4$, the simulation does not have galaxies with a specific SFR high enough to account for the observed starburst galaxies \citep{Caputi2017}. The mentioned discrepancies in the faint-end of the UV LF and SFR-M$^{*}$ plane can be due to either the absence of a particularly dust-poor galaxy population, which has been neither previously observed by \hers nor it has been included in the dwarf irregular galaxy population, or by a limitation on the set of templates. We analyse the impact of these discrepancies on our results in Appendix \ref{sec:SFRUV}.\par

In \spr{}, the spatial distribution of galaxies, which is fundamental for line intensity mapping, is included starting from the observed two-point correlation function using the algorithm by \citet{Soneira1978}. All the details of this procedure are reported in \citetalias{Bisigello2021}. Briefly, galaxies are distributed following an angular correlation function $w(\theta)=A_{w}\theta^{1-\gamma}$, with a power-law slope $\delta=\gamma-1 = 0.7$ as suggested by observations \citep[e.g.][]{Wang2013}. At the same time, the spatial correlation length $r_{0}$, of which A$_{w}$ represent the angular projection, has a dependence on stellar mass, as derived from observations \citep{Wake2011,Hatfield2016}:
\begin{equation}
\begin{array}{l}
 r_{0} \propto \begin{cases} M^{k_{M,1}}, & \mbox{if } M^{*}\leq M^{*}_{break} \\ M^{k_{M,2}}, & \mbox{if } M^{*}>M^{*}_{break}  \end{cases} \\
\end{array}
\end{equation}
with the stellar mass break log$_{10}({\rm M}^{*}_{break}/{\rm M}_{\odot})=$10.6, a low-mass slope of k$_{M,1}=$0.0959$\pm$0.0003 and a high-mass slope of k$_{M,2}=$0.181$\pm$0.006.  The full procedure is repeated splitting the mock catalogue on different redshift slices. \par
In the \spr{} workflow, the spatial distribution derived as a function of stellar mass, as just mentioned, is obtained after assigning physical properties to each simulated galaxies. Therefore, a reader, if preferred, can ignore the included method and match the \spr{} catalogue to a dark-matter simulation, using a stellar-to-halo mass relation \citep[e.g.][]{Girelli2020}.
\par

More information on \spr{} and its comparison with observations are available in \citetalias{Bisigello2021}.

\section{\cii and CO lines in \spr}\label{sec:lines}
In the next sections we summarise the relations considered to include the \cii and CO emission lines in \spr.
\subsection{\cii}\label{sec:cii_relations}
\begin{figure}
    \centering
    \includegraphics[width=0.98\linewidth,keepaspectratio]{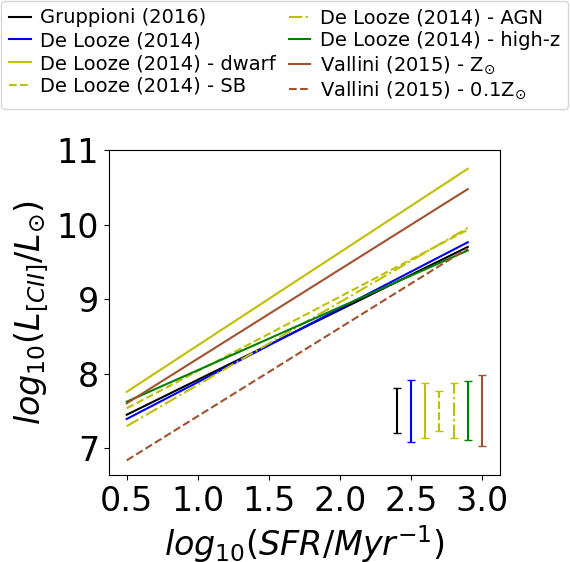}
    \caption{Comparison between the different relations used in this work to derive the \cii luminosity. For the relation by \citet{Gruppioni2016}, we convert L$_{IR}^{SF}$ to SFR using the relation by \citet{Kennicutt1998}. On the bottom right we show the uncertainties associated to each relation.}
    \label{fig:CII_relations}
\end{figure}

To include the \cii line in \spr, we assume three different methods to derive the expected \cii emission (Figure \ref{fig:CII_relations}). \par

First, we included the empirical relation derived by \citet{Gruppioni2016} starting from a local sample of Seyfert galaxies observed with \hers:
\begin{equation}
    \log(L_{[\ion{\rm C}{II}]}/{\rm L_{\odot}})=(0.94\pm0.03)\,\log(L^{SF}_{IR}/{\rm L_{\odot}})-(2.39\pm0.30)
\end{equation}
where L$^{SF}_{IR}$ is the component of the IR luminosity due to star-formation. In their work, they verify that this relation does not change with the AGN fraction, hence it can be used either in sources with no or little AGN contribution or in AGN dominated objects.\par
Second, we considered the empirical relation between SFR and L$_{[\ion{\rm C}{II}]}$ proposed by \citet{DeLooze2014} using a broad sample of galaxies, including dwarfs. In particular, they derived different relations for the overall galaxy sample and for different sub-samples of galaxies, i.e. metal-poor dwarf galaxies, star-forming or starburst galaxies, composite or AGN sources, and galaxies at high-\textit{z} ($z>0.5$). The relations are in the form of:
\begin{equation}\label{eq:D14}
    \log \left( \frac{\rm SFR}{{\rm M_{\odot} yr^{-1}}}\right)=\beta+\alpha\,\log(L_{[\ion{\rm C}{II}]}/{\rm L_{\odot}})
\end{equation}
whose $\alpha$ and $\beta$ depend on redshift and galaxy type \citep[see Table 3 in][]{DeLooze2014}.

\par
Finally, we included the relation by \citet[][ hereafter V15]{Vallini2015} derived by post-processing a radiative-transfer cosmological simulation at $z\sim6.6$. The post-processing was constructed to obtain on a cell-by-cell basis the \cii emission from dense PDRs and from the diffuse neutral gas \citep{Wolfire2003}. In particular, the \cii emission from \citetalias{Vallini2015} is well described by the following analytical relation, depending on metallicity (Z) and SFR:
\begin{equation}
\begin{split}
    \log&(L_{[\ion{\rm C}{II}]}/{\rm L_{\odot}})=\\
    &=7.0+1.2\log({\rm SFR})+0.021\log(Z/Z_{\odot})+\\
    &+0.012\log({\rm SFR})\log(Z/Z_{\odot})-0.74(\log(Z/Z_{\odot}))^2
\end{split}
\end{equation}
For this relation we consider a scatter similar to that derived by \citet{DeLooze2014} for star-forming galaxies, i.e. 0.27 dex, and a larger one of 0.37\footnote{Derived from the observed scatter of 0.48 dex and considering an error on the SFR of 0.3 dex.} dex for galaxies at z$>$5 \citep{Carniani2018}.

\subsection{CO lines}\label{sec:CO_relations}
The CO LF evolution at different $J$ is a great tool for shedding light on the underlying physical properties of different galaxy populations across cosmic time \citep[see e.g.][]{Obreschkow2009, Lagos2012, Vallini2016, Popping2019}. The relative luminosity of different CO lines, also referred to as CO Spectral Line Energy Distribution (CO SLED) brings in fact unique insights on the gas density and temperature, and on the heating mechanisms acting in the ISM of galaxies \citep[e.g.][]{Meijerink2007, Rosenberg2015, Pozzi2017, Mingozzi2018, Talia2018, Vallini2019}. The critical density of the CO rotational lines scales as $\propto J$, thus the low-$J$ lines ($J<4$) arise from diffuse cold ($n\approx 100-1000\rm \, cm^{-3}$, $T\approx 20-30 \rm K$) molecular gas -- predominantly heated by star formation creating PDRs \citep[][]{Hollenbach1999, Wolfire2022} -- while mid-$J$ ($4\leq J\leq 7$), and high-$J$ ($J>7$) CO lines trace increasingly dense and warm ($n> 10^4-10^5\rm \, cm^{-3}$, $T>50\rm \, K$) clumps within molecular clouds \citep[e.g.][]{McKee2007}. Mid and high-$J$ line luminosity can be enhanced by the X-ray heating from the AGN -- creating X-ray Dominated Regions \citep[XDRs; ][]{Maloney1996, Meijerink2007, Wolfire2022} -- and/or by shock heating from merging events or outflows \citep[e.g.][]{McKee1977}.\par

As for the \cii line, we considered different recipes to include CO $J\rightarrow(J-1)$ rotational transitions in \spr. We assume different CO excitation depending on the galaxy population, as summarised in Table \ref{tab:CO_references} and described in details in the next sections. In Appendix \ref{sec:co_sled} we show the CO SLEDs associated to each galaxy type using the different relations. The reader can refer to Figure \ref{fig:COSLED} to understand which relation consider to minimize discontinuities in the CO SLED. We assume that elliptical galaxies do not emit in CO, as few of them have been observed in CO even in the Local Universe \citep[$\sim22\%$;][]{Young2011} and, even trough stacking, they show, in general, a low gas fraction \citep[$<1\%$ in the Local Universe and $<8\%$ at $z\sim1.5$;][]{Magdis2021}. \par

\begin{table}
    \caption{Relations used to include CO lines in \spr{} for different $J$ and galaxy populations: a) \citet{Sargent2014}, b) \citet{Greve2014}, c) \citet{Boogaard2020}, d) \citet{Liu2015}, e) \citet{Rosenberg2015}, f) \citet{Liu2021}, g) \citet{Esposito2022}. For $J\geq14$ we considered as reference the observed CO SLED by \citet[][h]{Mashian2015}. }
        \centering
    \begin{tabular}{c|ccc}
       J  & Spirals, SF-AGN, & SB-AGN, AGN1, \\
          & SB, Irr & AGN2\\
       \hline
       1  & a & b \\
       2,3  & c,g & b,g \\
       4  & c,d,e,g & b,d,e,g \\
       5  & c,d,e,f,g & b,d,e,f,g \\
       6,7,8  & c,d,e,g & b,d,e,g \\
       9,10,11,12  & b,d,e,g & b,d,e,g \\
       13 & b,e,g & b,e,g \\
       $\geq$14 & h & h \\
    \end{tabular}
\label{tab:CO_references}
\end{table}

\subsubsection{$J<14$} 

In order to derive the CO $J<14$ luminosities\footnote{We use different notation depending on the unit of measurement of the CO luminosity. L$_{CO(1-0)}^{\prime}$ is measured in $\rm K\,km\,s^{-1}\,pc^{2}$, while L$_{CO(1-0)}$ is measured in L$_{\odot}$.}, we considered different relations from the literature, either based on the ratio of the CO $J\rightarrow(J-1)$ transitions with respect to the CO(1-0) or with respect to the \cii line. Moreover, we also tested relations between the IR and the CO luminosities. In particular, to derive the CO(1-0) of star-forming galaxies with or without a non-dominant AGN component (i.e. Spiral, SB, SF-AGN, Irr) we considered the relation by \citet{Sargent2014}:
\begin{equation}
    \log \left( \frac{L_{CO(1-0)}^{\prime}}{\rm K\,km\,s^{-1}\,pc^{2}} \right)=0.54+0.81\,\log \left( \frac{L_{IR}}{\rm L_{\odot}} \right).
\end{equation}
This was obtained in a sample of 131 galaxies at $z<3$ with ${\rm M}_*>10^{10}\,{\rm M}_{\odot}$, and has a scatter of 0.21 dex. \par
For the same galaxy populations, but from $J=2$ to $J=8$, we considered the recent $L_{CO(J\rightarrow(J-1))}^{\prime}/L_{CO(1-0)}^{\prime}$ ratios by \citet{Boogaard2020}, which were derived from observations of 22 star-forming galaxies up to $z=3.6$, as part of the ASPECS survey \citep{Walter2016}. The ratios, for $J$ from 2 to 6, are reported in Table \ref{tab:Bo20} for galaxies above and below $z=2$, except for $J=7$ and 8 for which a single value is present, corresponding to observations at $z>2$. To avoid discontinuities between $z<2$ and $z>2$, we interpolate between the two different ratios at $z=1.5-2.5$. 
To estimate the CO ratios, we used the CO(1-0) luminosity derived using the previously mentioned relation by \citet{Sargent2014}. \par

\begin{table}
    \caption{$L^{\prime}_{CO(J\rightarrow(J-1))}/L_{CO(1-0)}^{\prime}$ ratios by \citet{Boogaard2020} used for $J=2-8$ for spirals, SF-AGN, SB, and Irr. The first column indicates the $J$ value, the second and third columns the $L_{CO(J\rightarrow(J-1))}/L_{CO(1-0)}$ ratios at $z<2$ and $z>2$, respectively. For observational limitations, a single ratio is given for $J=7$ and 8 and it is used at all redshifts.}\label{tab:Bo20}    
    \centering
    \begin{tabular}{c|cc}
       $J$  &  \multicolumn{2}{c}{$L_{CO(J\rightarrow(J-1))}^{\prime}/L_{CO(1-0)}^{\prime}$}\\
          & $z<2$ & $z>2$ \\
          \hline
        2  & 0.75$\pm$0.11 & 0.97$\pm$0.15\\
        3  & 0.46$\pm$0.09 & 0.80$\pm$0.14\\
        4  & 0.25$\pm$0.07 & 0.61$\pm$0.13\\
        5  & 0.12$\pm$0.06 & 0.44$\pm$0.11\\
        6  & 0.04$\pm$0.05 & 0.28$\pm$0.09\\
        7  & \multicolumn{2}{c}{0.17$\pm$0.07} \\
        8  & \multicolumn{2}{c}{0.09$\pm$0.06} \\
    \end{tabular}

\end{table}

Some studies have shown that the CO(1-0)-SFR relation is different for low-metallicity galaxies \citep[e.g;][]{Cormier2014}. Therefore, we applied a correction to the CO(1-0) luminosities for galaxies with sub-solar metallicity (i.e. 12+log$_{10}(O/H)<$8.7), following the results derived by \citet{Hunt2015}:
\begin{equation}
\begin{split}
    \log&({\rm SFR}/L_{CO(1-0)}^{\prime})=\\
    &(-2.25\pm0.15)[12+\log(O/H)]+(11.31\pm1.3)
\end{split}
\end{equation}
\par
For the galaxy populations with a dominant AGN component (i.e., SB-AGN, AGN1 and AGN2) we considered the relation between CO(1-0) and total IR luminosity by \citet{Greve2014}:
\begin{equation}
    \log \left( \frac{L_{IR}}{\rm L_{\odot}} \right)=2.00\pm0.5+(1.00\pm0.05)\,\log \left( \frac{L_{CO(1-0)}^{\prime}}{{\rm K\,km\,s^{-1}\,pc^{2}}} \right).\label{eq:CO1_0_AGN}
\end{equation}
with a scatter of 0.27 dex. The relation was derived from a sample of 62 local ultra-luminous infrared galaxies (ULIRGs), but consistent results were obtained including AGN-dominated systems \citep{Greve2014}. We also included the relations presented in the same paper to convert the total IR luminosity to the luminosity of the $J=2$--13 CO transitions. \par
We explored also another possibility for deriving the  $1<J<14$ transitions for the same galaxy populations, namely we considered the CO($J\rightarrow(J-1)$)/CO(1-0) ratios from a sample of 35 local AGNs \citep[$\rm L_{2-10keV} \geq 10^{42}\, erg\,s^{-1}$;][]{Esposito2022}. In particular, we derived the CO($J\rightarrow(J-1)$)/CO(1-0) ratios after cross-matching the sample by \citet{Esposito2022} with that by \citet{Gruppioni2016} to identify galaxies with an AGN fraction at 5-40 $\mu$m above and below 40$\%$ (8 and 7 objects, respectively). This threshold on the AGN fraction separates, in \spr, the SF-AGN ($f_{AGN}<40\%$) to the other AGN populations (i.e. SB-AGN, AGN1 and AGN2). We then derived the median CO($J\rightarrow(J-1)$)/CO(1-0) ratios for both AGN populations (Table \ref{tab:Es22}) and we normalised such ratios to the CO(1-0) derived in Eq. \ref{eq:CO1_0_AGN}. 

\begin{table}
    \centering
    \caption{Median $L_{CO(J\rightarrow(J-1))}^{\prime}/L_{CO(1-0)}^{\prime}$ ratios for objects with an AGN fraction at 5-40$\mu$m above and below 40\%. Values are derived by cross-matching the samples by \citet{Esposito2022} and \citet{Gruppioni2016}.}
    \begin{tabular}{c|cc}
       $J$  &  \multicolumn{2}{c}{$L_{CO(J\rightarrow(J-1))}^{\prime}/L_{CO(1-0)}^{\prime}$}\\
          & f$_{AGN}$(5-40$\mu$m)>40\% & f$_{AGN}$(5-40$\mu$m)<40\% \\
          \hline
    2 & 0.897$_{-0.543 }^{+ 0.633 }$ & 0.437$_{- 0.134 }^{+ 0.428 }$ \\
    3 & 0.819$_{0.644 }^{+ 0.691 }$ & 0.457$_{- 0.158 }^{+ 0.533 }$ \\
    4 & 0.964$_{-1.078 }^{+ 0.981 }$ & 0.266$_{- 0.181 }^{+ 0.300 }$ \\
    5 & 0.404$_{-0.379 }^{+ 0.329 }$ & 0.161$_{- 0.049 }^{+ 0.142 }$\\
    6 & 0.154$_{-0.083 }^{+ 0.138 }$ & 0.097$_{- 0.030 }^{+ 0.090 }$\\
    7 & 0.030$_{-0.017 }^{+ 0.064 }$ & 0.052$_{- 0.015 }^{+ 0.049 }$\\
    8 & 0.049$_{-0.033 }^{+ 0.034 }$ & 0.030$_{- 0.009 }^{+ 0.026 }$ \\
    9 & 0.080$_{-0.065 }^{+ 0.063 }$ & 0.014$_{- 0.075 }^{+ 0.013 }$ \\
    10 & 0.024$_{-0.014 }^{+ 0.023 }$ & 0.010$_{- 0.012 }^{+ 0.010 }$\\
\end{tabular}
    \label{tab:Es22}
\end{table}

\par
In \citet{Liu2015}, the FIR (40-400 $\mu$m) luminosity of 167 local galaxies with \hers spectroscopic observations is related to the CO luminosity from $J=4$ to $J=12$, as:
\begin{equation}\label{eq:CO_L15}
    \log\left(\frac{L_{FIR}}{{\rm L}_{\odot}}\right)=N\,\log\left(\frac{L_{CO(J\rightarrow(J-1))}^{\prime}}{\rm K\,km\,s^{-1}\,pc^{2}}\right)-A
\end{equation}
with the values of N and A reported in Table \ref{tab:CO_L15}. In \spr{} we applied these relations to each simulated galaxy, without separating for galaxy population and redshift. We note that beyond $J=10$ no observed LFs are available for comparison. \par

\begin{table}
    \caption{Slope and intercept of the $L_{FIR}-L_{CO(J\rightarrow(J-1))}^{\prime}$ relations (see. eq. \ref{eq:CO_L15}) taken from \citet{Liu2015}. The last column show the 1$\sigma$ scatter around the relations.}
        \centering
    \begin{tabular}{c|ccc}
        $J$ & N & A & $\sigma$ [dex] \\
        \hline
        4 & 1.06$\pm$0.03 & 1.49$\pm$0.24 & 0.265\\
        5 & 1.07$\pm$0.03 & 1.71$\pm$0.22 & 0.218\\
        6 & 1.10$\pm$0.03 & 1.79$\pm$0.24 & 0.192\\
        7 & 1.03$\pm$0.04 & 2.62$\pm$0.26 & 0.193\\
        8 & 1.02$\pm$0.03 & 2.82$\pm$0.27 & 0.210\\
        9 & 1.01$\pm$0.03 & 3.10$\pm$0.22 & 0.267\\
        10 & 0.96$\pm$0.04 & 3.67$\pm$0.25 & 0.259\\
    \end{tabular}
    \label{tab:CO_L15}
\end{table}

For $J\geq4$, we included the CO luminosities derived considering the CO($J\rightarrow(J-1)$)/\cii ratio from \citet{Rosenberg2015} and considering the three classes of objects presented on the same paper. In particular, the first class (c1) does not require any mechanisms in addition to the UV-heating from star-formation to reproduce the observed CO-ladder, while the third class (c3) includes galaxies with an AGN-component and probably requires mechanical heating in addition to UV-heating to describe its excited CO-ladder. The second class (c2) simply indicates a intermediate case, where it is not possible to discriminate which heating mechanism dominate the CO-ladder. The CO($J\rightarrow(J-1)$)/\cii ratio should be more stable than CO($J\rightarrow(J-1)$)/CO(1-0) given the faintness of CO(1-0) lines. \par
The ratios for the different classes are reported in Table \ref{tab:CO_CII_R15}. In this work we first considered the three extreme cases where all galaxies behave like a single class, then we examined the case where spiral, SB and Irr are in c1, SF-AGN are in c2, while SB-AGN, AGN1 and AGN2 are in c3. We fixed as ground values for the \cii luminosities those derived considering the relation by \citet[][see Sec. \ref{sec:cii_relations}]{Vallini2015}, given its agreement with the observed \cii LF (see \ref{sec:cii_obs}).\par

\begin{table}
    \caption{CO($J\rightarrow(J-1)$)/\cii luminosity ratio for the three classes derived from \citet{Rosenberg2015}.}
    \centering
    \begin{tabular}{c|ccc}
    $J$ & \multicolumn{3}{c}{CO($J\rightarrow(J-1)$)/\cii} \\
    & c1 & c2 & c3 \\
    \hline
4 &	0.0158$\pm$0.0263 & 0.0387$\pm$0.0480 &	0.0295$\pm$0.0911 \\
5 &	0.0202$\pm$0.0303 & 0.0459$\pm$0.0462 &	0.0298$\pm$0.1031 \\
6 & 0.0190$\pm$0.0296 &	0.0545$\pm$0.0589 &	0.0399$\pm$0.1318 \\
7 & 0.0170$\pm$0.0267 &	0.0559$\pm$0.0631 &	0.047$\pm$0.1536 \\
8 &	0.0141$\pm$0.0228 &	0.0521$\pm$0.0666 &	0.0553$\pm$0.1923 \\
9 &	0.0128$\pm$0.0147 &	0.0449$\pm$0.0604 &	0.0441$\pm$0.1939 \\
10 & 0.0085$\pm$0.0106 & 0.0393$\pm$0.0420 & 0.0372$\pm$0.2170 \\
    \end{tabular}
    \label{tab:CO_CII_R15}
\end{table}

Finally, \citet{Liu2021} focused on the $R_{52}=$CO(5-4)/(2-1) line ratio, showing its dependence on the total IR luminosity:
\begin{equation}
    R_{52}=0.18\,\log(L_{IR}/{\rm L}_{\odot})-1.83
\end{equation}
We included also this relation in \spr{} with a scatter of 0.18 dex, as reported in the reference paper, without distinguishing between different galaxy populations or redshifts.\par

\subsubsection{$J\geq$14} 
At present, no relations are available in the literature to derive the CO transitions with $J\geq14$ starting from the SFR or the IR luminosity. Therefore, we decided to adopt observed ratios as reference for our simulated galaxies. In particular, we considered the CO($J\rightarrow(J-1)$)/CO(1-0) (Table \ref{tab:J14}) estimated from observations by \citet{Mashian2015}, using NGC6240 as a reference for SB-AGN, Mrk231 for AGN1 and AGN2, and M82 for SB. These galaxies are also among the templates included in \spr{} to derive photometry and physical properties for the same galaxy populations. \par
The CO SLED of NGC6240 is detected up to $J=24$, while Mrk231 and M82 are detected up at $J=20$ and $J=18$, respectively. Beyond these transitions, only upper limits are available and we therefore considered no CO emission. The CO SLED of Mrk231 has been previously studied \citep{vanderwerf2010,Vallini2019}, showing that the excitation of the CO $J>8$ lines cannot be completely reproduced considering the PDR emission only, but it requires also an XDR component created by the X-rays from the accretion onto the central black hole. Moreover, the CO emissions for $J\geq13$ are completely dominated by the emission coming from the XDR. Given the absence of such a source of high X-ray excitation in the spiral, SF-AGN and dwarf populations,  we assumed that their CO $J\geq14$ transitions are negligible.

\begin{table}
    \caption{$L_{CO(J\rightarrow(J-1))}^{\prime}/L_{CO(1-0)}^{\prime}$ ratios considered for SB, SB-AGN and, AGN1 and AGN2, as taken from observations by \citet{Mashian2015}.}
    \centering
    \begin{tabular}{c|ccc}
         $J$ & SB & SB-AGN & AGN1\&AGN2 \\
           & (M82) & (NGC6240) & (Mrk231) \\
          \hline
         14 & 0.0047$\pm$0.0005$^{a}$ & 0.0311$\pm$0.0070 & 0.0049$\pm$0.0013\\
         15 & 0.0023$\pm$0.0005 & 0.0279$\pm$0.0063& 0.0100$\pm$0.0022\\
         16 & 0.0009$\pm$0.0002& 0.0234$\pm$30.0074$^{a}$& 0.0049$\pm$0.0009\\
         17 & 0.0006$\pm$0.0002$^{a}$& 0.0199$\pm$0.0044& 0.0058$\pm$0.0017$^{a}$\\
         18 & 0.0004$\pm$0.0001& 0.0072$\pm$0.0015 & 0.0063$\pm$00013\\
         19 & -- & 0.0060$\pm$0.0013& 0.0039$\pm$0.0012$^{a}$ \\
         20 & -- & 0.0036$\pm$0.0007 & 0.0021$\pm$0.0005\\
         21 & -- & 0.0039$\pm$0.0008 & -- \\
         22 & -- & 0.0039$\pm$0.0008& -- \\
         23 & -- & 0.0041$\pm$0.0009& -- \\
         24 & -- & 0.0030$\pm$0.0006& --\\
    \end{tabular}
    \tablefoot{
    \tablefoottext{a}{We calculated this value averaging the two ratios at $J-1$ and $J+1$.}
    }    
    \label{tab:J14}
\end{table}

\section{Comparison with CO and \cii observations}\label{sec:comp_obs}
In this section we compare some observed \cii and CO LFs with the LFs in \spr{} derived considering all the relations previously discussed. All LFs are made publicly available\footnote{\url{http://spritz.oas.inaf.it/}\label{spriztsite}}. We also report the $1\sigma$ confidence intervals associated to each LFs. These intervals are obtained taking into account the uncertainties associated to the observed LFs or GSMFs used as inputs in the simulation (see Sec. \ref{sec:SPRITZ}) and the errors associated to the relations used to derive the \cii or CO line luminosities. The LFs are derived considering the full \spr{} catalogue (i.e., $L_{IR}>10^{6}L_{\odot}$) and the entire sky, as the observed LFs considered in the next sections for the comparison are corrected for incompleteness and include cosmic variance in their errors.

\begin{figure*}
    \centering
    \includegraphics[width=0.98\linewidth,keepaspectratio]{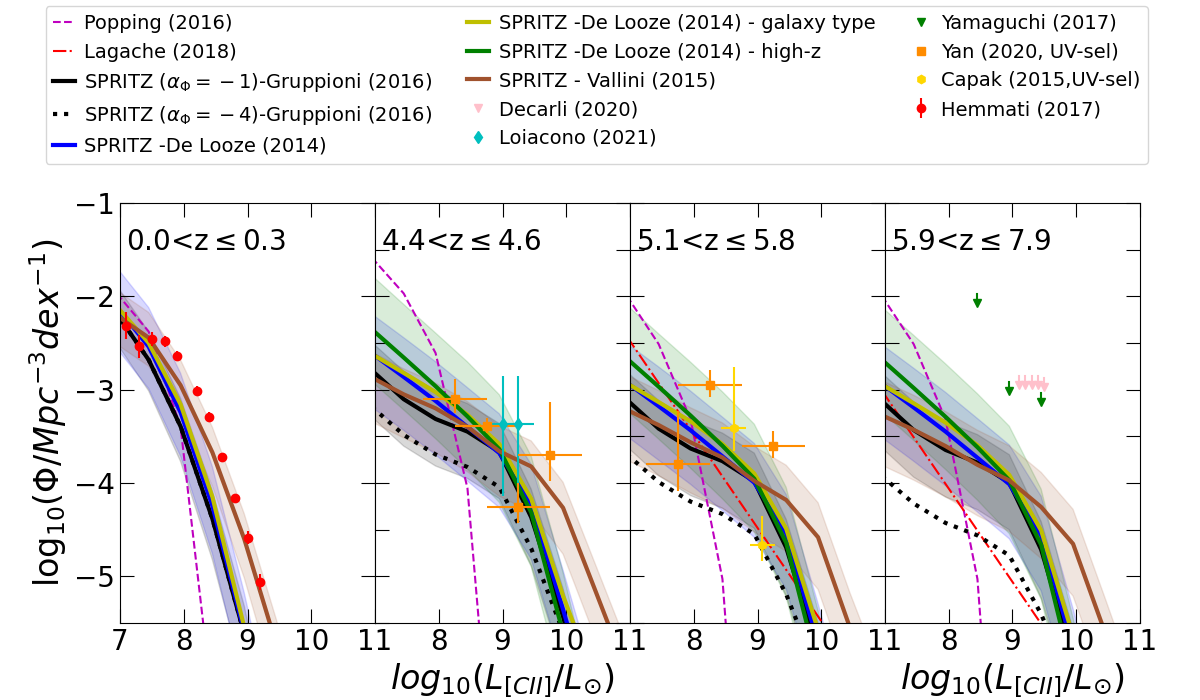}
    \caption{\cii LF derived with \spr{} assuming different relations (see Section \ref{sec:cii_relations}). We compare the results with \cii observations by \citet{Hemmati2017} in the Local Universe, observations at z$\sim$4.45 by \citet{Yan2020} and \citet{Loiacono2021}, the ones at $z\sim$5.5 by \cite{Capak2015}, at $z\sim$6.3 by \cite{Yamaguchi2017} and at $z\sim$6.9 \cite{Decarli2020}. We also report the model predictions of \citet[][\textit{magenta dashed lines}]{Popping2016} and \citet[][\textit{red dotted lines}]{Lagache2018} at $z\geq4.7$. The shaded areas (same colours as the solid lines) show the uncertainties of the considered \cii relations and the 1$\sigma$ errors on the \spr{} input LFs and GSMFs. }
    \label{fig:LCII}
\end{figure*}

\begin{figure*}
    \centering
    \includegraphics[width=0.98\linewidth,keepaspectratio]{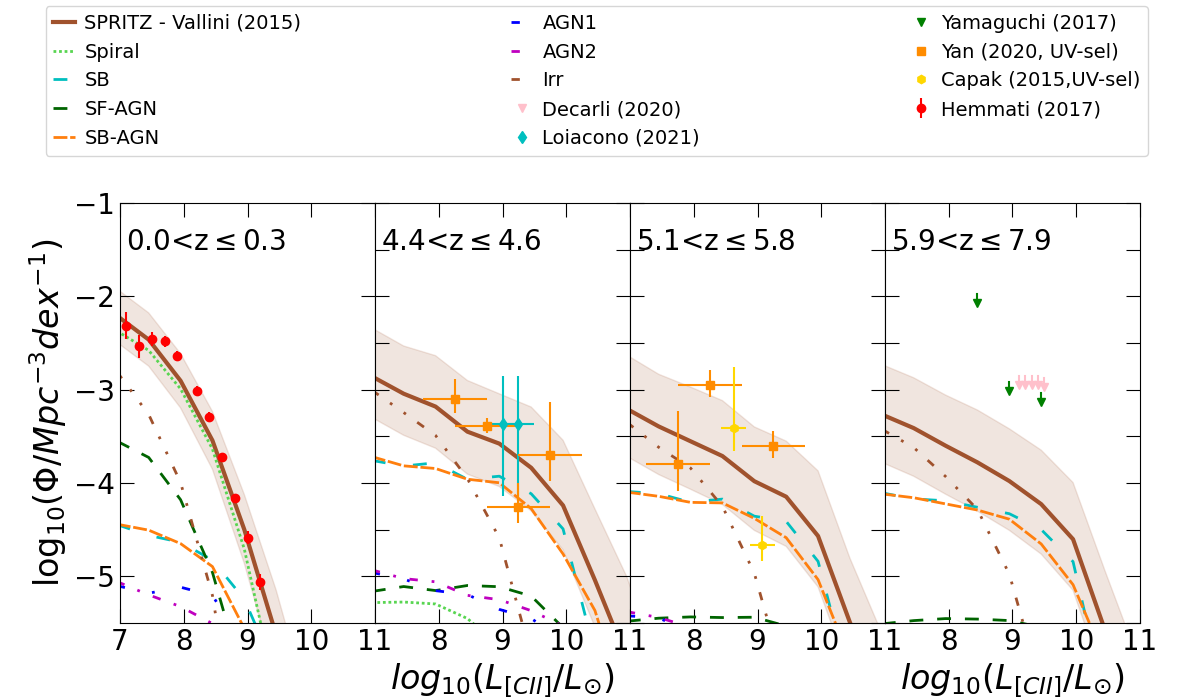}
    \caption{The same as Figure \ref{fig:LCII}, but we report only the relation by \citetalias{Vallini2015} (\textit{solid brown line}) and we separate the LF by the different galaxy populations (\textit{coloured lines}): spirals (\textit{dotted light green}), SB (\textit{loosely dashed cyan}), SF-AGN (\textit{dashed dark green}), SB-AGN (\textit{densily dashed orange}), AGN1 (\textit{loosely dash-dotted blue}), AGN2 (\textit{dash-dotted magenta}) and Dwarf irregulars (\textit{dash-dot-dotted brown}).}
    \label{fig:LCII_SED}
\end{figure*}

\subsection{\cii}\label{sec:cii_obs}

In Figure \ref{fig:LCII}, we report the \spr{} \cii LF compared with observational results by \cite{Capak2015,Yamaguchi2017,Hemmati2017,Decarli2020,Yan2020} and \citet{Loiacono2021}. Results by \citet{Capak2015,Hemmati2017} and \citet{Yan2020} are corrected for incompleteness, while such correction in the work by \citet{Loiacono2021} is limited, given that it is based on a single detection. The works by \citet{Hemmati2017} and \citet{Loiacono2021} are based on direct \cii observations, based on a blind \hers survey and on a ALMA serendipitous detection, respectively. On the other hand, the results by \citet{Capak2015} and \citet{Yan2020} are derived from a sample of UV-selected galaxies: for this reason, they may be affected by observational biases. \par
Given the absence of spectroscopic instruments covering the wavelength range between 160 and 600 $\mu$m, no observed \cii LFs are currently available to fill the gap between the Local Universe and $z\sim4.5$. In the same Figure we also report, for a direct comparison with our predictions, the \cii LF derived from the semi-analytical models by \citet{Popping2016} and \citet{Lagache2018}, which is valid at $4.7\leq z \leq 8$.\par
In the Local Universe, \spr{} can reproduce the observed values only when considering the relation by \citetalias{Vallini2015}, which includes a dependence on both SFR and metallicity. It is worth noticing that the relation by \citetalias{Vallini2015} is in agreement with the relation by \citet{DeLooze2014} for solar metallicity, once considered the respective uncertainties. However, the relation by \citetalias{Vallini2015} has a steeper slope (i.e. log($\rm L_{[\ion{\rm C}{II}]}) \propto1.2\,{\rm log(SFR)}$) that the relation by \citet[][i.e. $\rm log( L_{[\ion{\rm C}{II}]}) \propto1.0\,{\rm log(SFR)}$; see Figure \ref{fig:CII_relations}]{DeLooze2014}, which is responsible for the better agreement in the Local Universe, where metallicity has a minor impact. Moreover, the need for a steeper slope in the \cii-SFR is not driven by the underestimation of the UV-component of the SFR in \spr, as it has a negligible effect at low-$z$ (see Appendix \ref{sec:SFRUV}).  \par
At z$>$4 the observed values show a significant dispersion and all the relations, which mainly differ at L$_{[\ion{\rm C}{II}]}>10^{9.5}$ L$_{\odot}$, are broadly consistent with the observations. The LFs reported in the Figure correspond to the flattest high-\textit{z} extrapolation (i.e. $\Phi^{*}\propto(1 +\textit{z})^{-1}$) included in \spr, but we also report, as an example, the LF derived considering the relation by \citet{Gruppioni2016} and a number density at the knee ($\Phi^{*}$) decreasing as $\propto(1 +\textit{z})^{-4}$. The latter LF is well below the observed values, showing that the data are consistent with the first extrapolation (i.e. $\Phi^{*}\propto(1 +\textit{z})^{-1}$), as also observed for the total IR LFs \citepalias{Bisigello2021}. \par

\begin{figure}
    \centering
    \includegraphics[width=\linewidth,keepaspectratio]{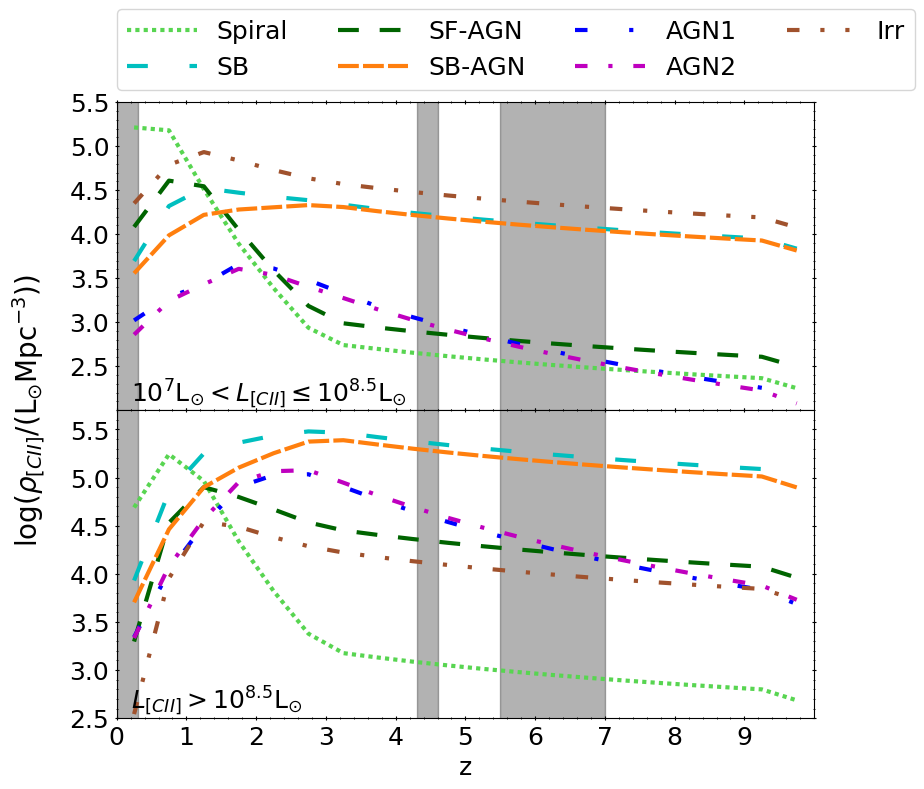}
    \caption{Contribution to the \cii luminosity density of the different galaxy populations (see legend) included in \spr. We report two different luminosity ranges: $10^{7}\,\rm{L}_{\odot}<L_{[\ion{\rm C}{II}]}\leq10^{8.5}\,\rm{L}_{\odot}$ (\textit{top}) and $L_{[\ion{\rm C}{II}]}>10^{8.5}\,\rm{L}_{\odot}$ (\textit{bottom}). The grey shaded areas show the redshift range where observations are available.  We highlight that SB and SB-AGN have sSFR ranging from $\rm log(\rm sSFR/yr^{-1})=-8.8$ to -8.1, which is above the the main sequence at low-z, but on it at $z=5-7$.}
    \label{fig:LCII_SED_z}
\end{figure}

In Figure \ref{fig:LCII_SED_z}, we split the \cii LF, derived considering the relation by \citetalias{Vallini2015}, into the different contributions of the single galaxy populations. In this way, we can appreciate that in the Local Universe the \cii LF is dominated at all luminosities by spiral galaxies, while at z$>$4 it is dominated by dwarf irregular galaxies at L$_{[\ion{\rm C}{II}]}<10^{8.5}$ L$_{\odot}$ and by SB and SB-AGN at brighter luminosities. These two populations have sSFR ranging from $\rm log(\rm sSFR/yr^{-1})=-8.8$ to -8.1, therefore, they are considerably above the main sequence at low-z, but they are in the main sequence at $z=5-7$. Indeed, following the parametrisation by \citet{Speagle2014}, the main sequence at $z=5$ corresponds to a $\rm log(\rm sSFR/yr^{-1})=-8.5$ to -7.9, depending on the stellar mass, with an observed scatter of 0.3 dex. This is also consistent with the results by \citet{Faisst2020}, which found that ALPINE sources at $z=$4-6 are star-forming galaxies on the main sequence. We remind the reader that, as explained in Section \ref{sec:SPRITZ}, the templates associated to each galaxy population do not evolve with redshift. \par
Going more into details, in \spr{} the faint-end of the \cii LF moves from being dominated by spiral to being dominated by dwarf irregulars around $z\sim1$ (Figure \ref{fig:LCII_SED_z}). On the other hand, the contribution of the SB and the SB-AGN populations to the bright-end of the \cii LF becomes dominant at $z\sim1$ and $z\sim2.8$. It is however evident the need of more observations between $z=0.3$--4 to verify these predictions. \par
Given the urgency for future far-IR probes covering the gap between \hers ($z=0$) and ALMA ($z>4$) \cii observations, we report for reference in Appendix \ref{sec:cii_allz} the \cii LF derived by \spr, considering the relation by \citetalias{Vallini2015}, up to $z=10$. Similarly, we also report the area and depth necessary to sample the \cii LF in different luminosity regimes. 

\subsection{CO lines}\label{sec:COobs}
We now compare the observations by \citet{Saintonge2017}, \citet{Riechers2019} and \citet{Decarli2019,Decarli2020}, all corrected for incompleteness in the respective works, with the different relations (see Table \ref{tab:CO_references}) considered in \spr{} to estimate CO luminosities. The work by \citet{Saintonge2017} is based on a representative sample of galaxies extracted starting from a mass-selected sample of galaxy ($\rm M_{*}>10^{9}\rm M_{\odot}$) at $z<0.05$, while the other two works are based on blind line observations. In this section we analyse only the CO transitions with $J\leq10$, as no observed LFs are available for higher J-values. To facilitate the comparison between the different models, we report in Table \ref{tab:chi2_CO} the total $\chi^{2}$ derived comparing each model with the available observations\footnote{The number of observed data points used to calculate the $\chi^{2}$ is the same for all models at fixed J value and redshift.}, taking into account the observational errors. 

\begin{table*}[]
    \centering
    \caption{$\chi^{2}$ derived comparing the different models with CO observations by \citet{Saintonge2017}, \citet{Riechers2019} and \citet{Decarli2019,Decarli2020}. Gr14, Bo20, Es22, Liu15, Liu21 and R15 refers to the relations by \citet{Greve2014}, \citet{Boogaard2020}, \citet{Esposito2022}, \citet{Liu2015}, \citet{Liu2021} and \citet{Rosenberg2015}, respectively. The last column (N$_{obs}$) indicates the number of observed data points, not always independent, used to calculate the $\chi^{2}$. }
        \begin{tabular}{cc|ccccccccc}
        J & $<z>$ & \multicolumn{8}{c}{$\chi^{2}$} \\
        & & Gr14\&Bo20 & Es22\&Bo20 & Liu15 & Liu21 & R15-c1 & R15-c2 & R15-c3 & R15-comb & N$_{obs}$ \\
    \hline
        1 & 0.15 & 9.7 &  -- & -- & -- & -- & -- & -- & -- & 19 \\
        1 & 2.25 & 12.64 & -- & -- & -- & -- & -- & -- & -- & 16 \\
        2 & 1.40 & 22.04 & 23.01 & -- & -- & -- & -- & -- & -- & 11 \\
        2 & 5.80 & 1.43 & 0.92 & -- & -- & -- & -- & -- & -- & 1 \\
        3 & 0.50 & 0.24 & 0.24 & -- & -- & -- & -- & -- & -- & 5\\
        3 & 2.60 & 3.25 & 2.24 & -- & -- & -- & -- & -- & -- & 11\\
        4 & 1.00 & 5.98 & 5.43 & 5.99 & -- & 3.38 & 0.55 & 1.00 & 2.30 & 8 \\
        4 & 3.70 & 16.12 & 14.86 & 22.33 & -- & 21.21 & 13.70 & 16.04 & 19.16 & 5 \\
        5 & 1.40 & 24.63 & 19.78 & 20.73 & 2.34 & 9.33 & 3.46 & 6.02 & 7.24 & 7 \\
        6 & 1.90 & 11.86 & 13.66 & 23.91 & -- &  8.09 & 2.82 & 3.91 & 6.14 & 7\\
        7 & 2.35 & 3.55 & 6.44 & 14.01 & -- & 3.93 & 0.89 & 1.15 & 2.33 & 6\\
        8 & 2.85 & 2.39 & 5.31 & 8.89 & -- & 2.43 & 0.25 & 0.23 & 0.78 & 5\\
        9 & 3.35 & 4.81 & 3.27 & 7.06 & -- & 3.05 & 0.69 & 0.68 & 1.50 & 3\\
        10 & 3.80 & 7.19 & 11.06 & 10.71 & -- & 7.61 & 1.08 & 1.26 & 2.98 & 3\\
    \end{tabular}
\label{tab:chi2_CO}
\end{table*}

\subsubsection{low-$J$}
\begin{figure}
    \centering
    \includegraphics[width=0.98\linewidth,keepaspectratio]{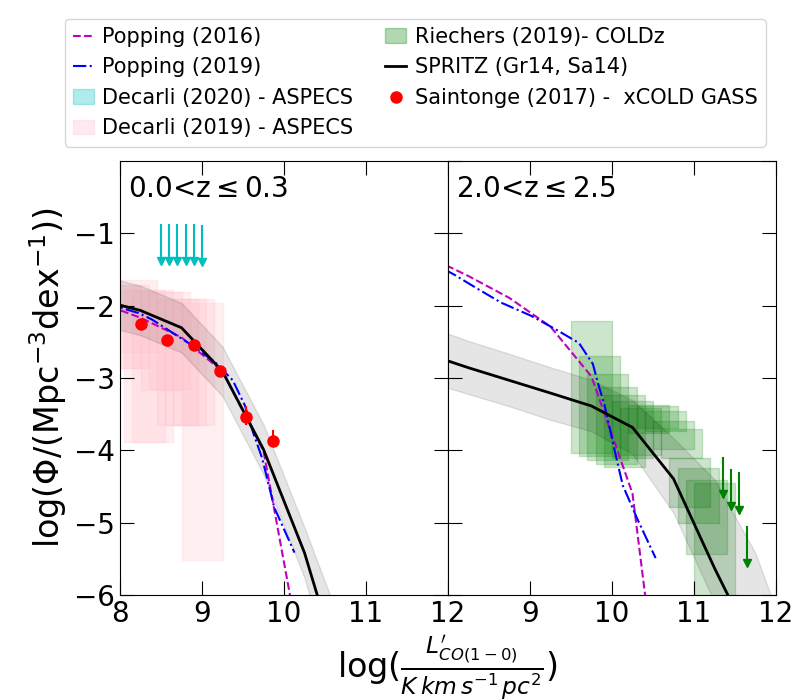}
    \caption{CO(1-0) LF (\textit{black solid lines}) derived with \spr{} compared with observations by \citet[\textit{red circles}][]{Saintonge2017}, \citet[][\textit{green squares}]{Riechers2019}, \citet[][\textit{pink squares}]{Decarli2019}, and \citet[][\textit{cyan arrows as upper limits}]{Decarli2020}. The grey shaded area contains the propagated errors of the IR LF and errors in the relations used to derive the CO(1-0) luminosity. We also report the models by \citet[][\textit{magenta dashed lines}]{Popping2016} and \citet[][\textit{blue dash-dotted line}]{Popping2019b}. Gr14 and Sa14 refer to the relations by \citet{Greve2014} and \citet{Sargent2014}, respectively.}
    \label{fig:LCO_1}
\end{figure}

\begin{figure}
    \centering
    \includegraphics[width=0.98\linewidth,keepaspectratio]{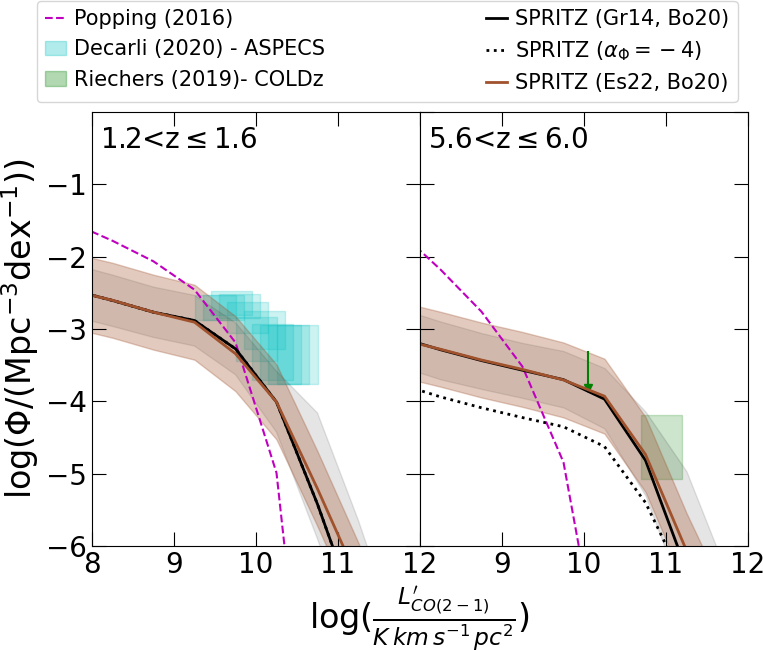}
    \includegraphics[trim={0 0 0 2.cm},clip,width=0.98\linewidth,keepaspectratio]{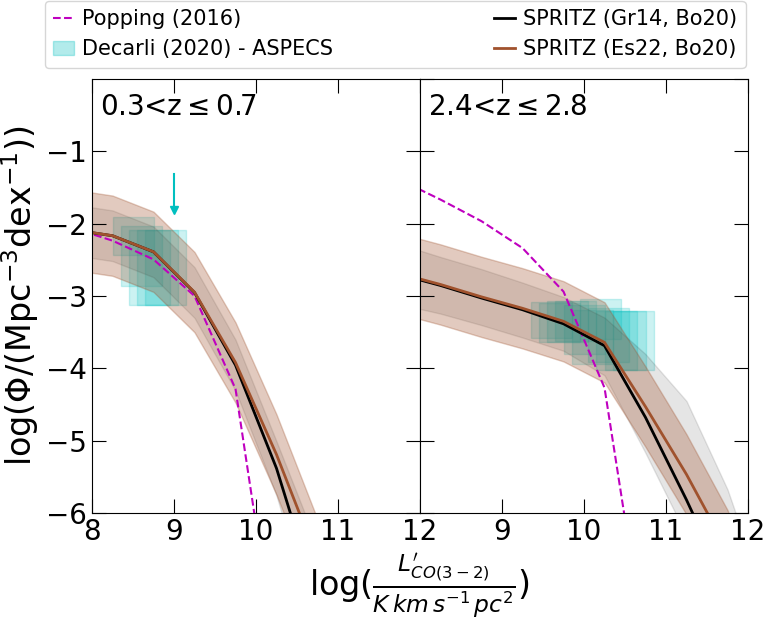}
    \caption{The same as Figure \ref{fig:LCO_1}, but for CO(2-1) (\textit{top}) and CO(3-2) (\textit{bottom}). The black dotted line present at $z>3$ indicates the CO(2-1) LF derived in \spr{} considering the steepest extrapolation considered for the number density (i.e. $\Phi^{*}\propto(1 +\textit{z})^{-4}$). Bo20 and Es22 refer to the relations by \citet{Boogaard2020} and \citet{Esposito2022}, respectively.}
    \label{fig:LCO_2_3}
\end{figure}

In Figures \ref{fig:LCO_1} and \ref{fig:LCO_2_3} we report the results for ${J}=1$ to 3. The \spr{} LFs are in agreement with the observations for these $J$ values at the available redshifts, with no strong difference between the results derived with the different relations ($\Delta\chi^{2}\leq1$). The main discrepancy between models and observations is for ${J}=2$ at $z\sim1.4$, where the \spr{} LFs are lower than the observed LFs by $\sim$0.5 dex, but they are still consistent within the errors. This difference is possibly linked to the light offset present at $z=1.0-1.7$ between the parametric description of the redshift evolution of the IR LF included as input in \spr{} and the observed \hers IR LF (see Appendix \ref{sec:LIR}). As for the \cii LF, the slope of the density extrapolation $\Phi^{*}\propto(1 +\textit{z})^{-4}$ in \spr{} is discarded by the data, as it leads to a significant underestimation (i.e. up to 0.8 dex) of the observed ${J}=2$ LF at $z\simeq6$ (see black dotted line in Figure \ref{fig:LCO_2_3}).\par
In the same Figures \ref{fig:LCO_1} and \ref{fig:LCO_2_3}, we also report the models by \citet{Popping2016,Popping2019}, which are close to our predictions for the CO(1-0) at $z<0.3$ and CO(3-2) at $z\sim0.5$. However, at $z>2$ the agreement between the data and our predictions is much better than with those of the \citet{Popping2016,Popping2019} models, particularly at the bright end. To investigate more this discrepancy, it would be interesting to verify if those models are also missing, or at least underpredicting, the most massive and dusty galaxies and, therefore, underestimating the bright-end of the IR LF. 

\subsubsection{Mid- and high-$J$}

\begin{figure*}
    \centering
    \includegraphics[width=0.45\linewidth,keepaspectratio]{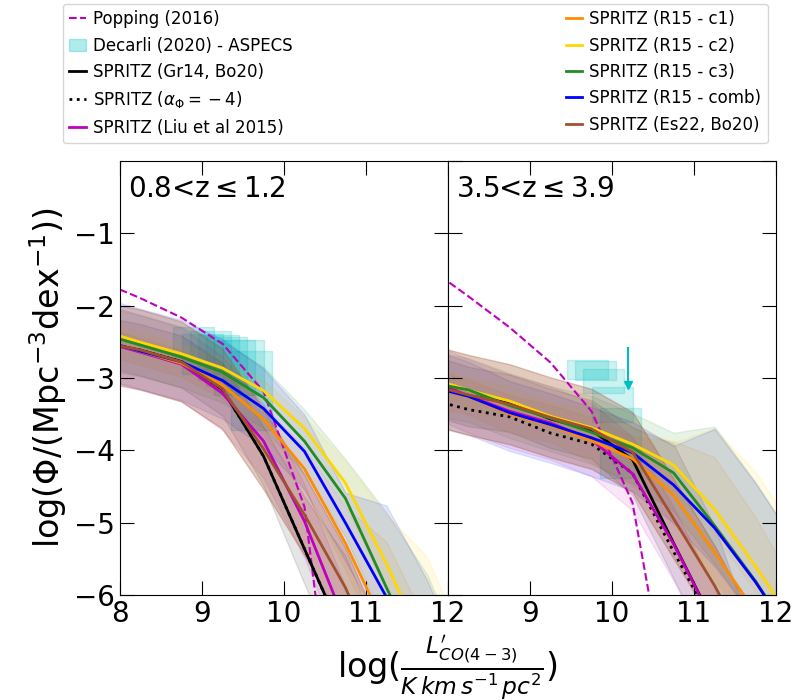}
    \includegraphics[width=0.45\linewidth,keepaspectratio]{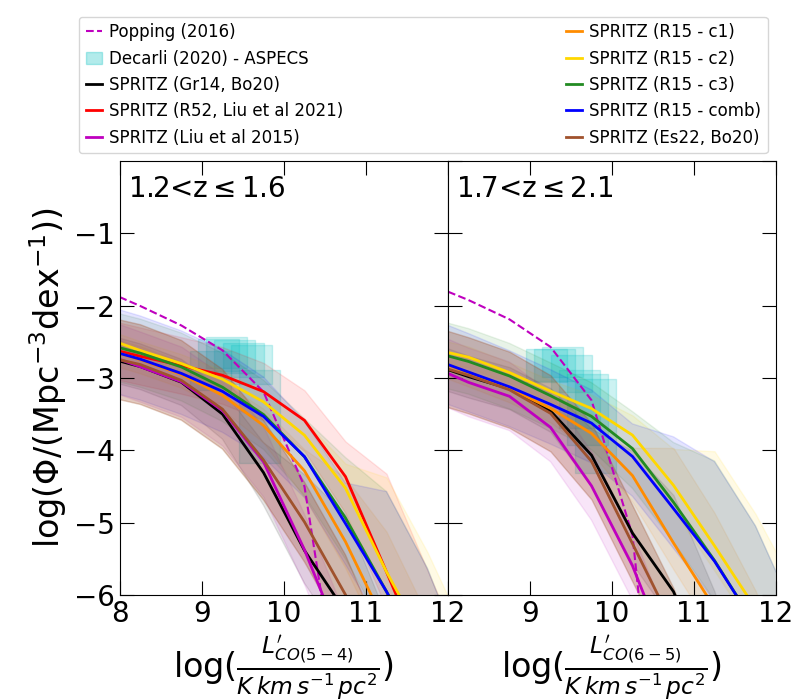}
    \includegraphics[width=0.45\linewidth,keepaspectratio]{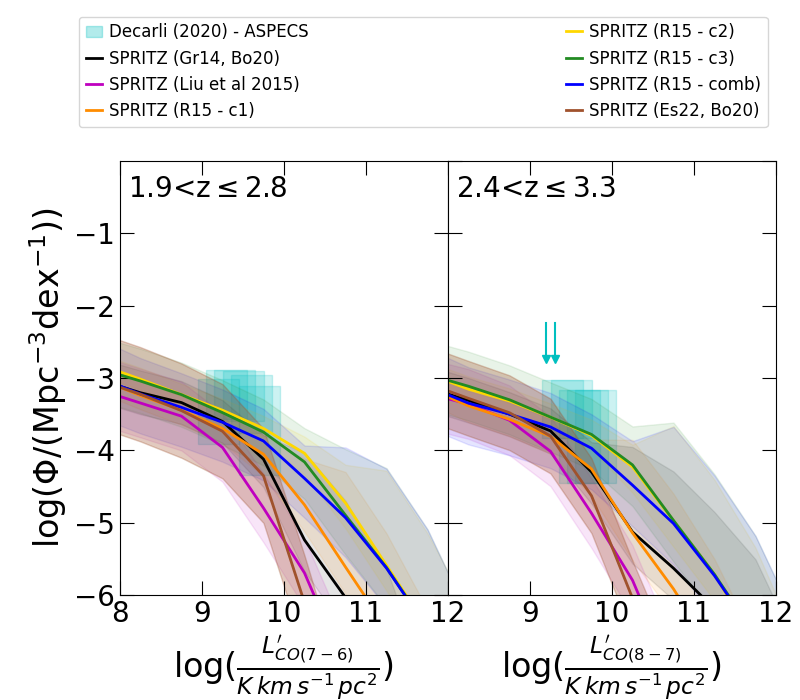}
    \includegraphics[width=0.45\linewidth,keepaspectratio]{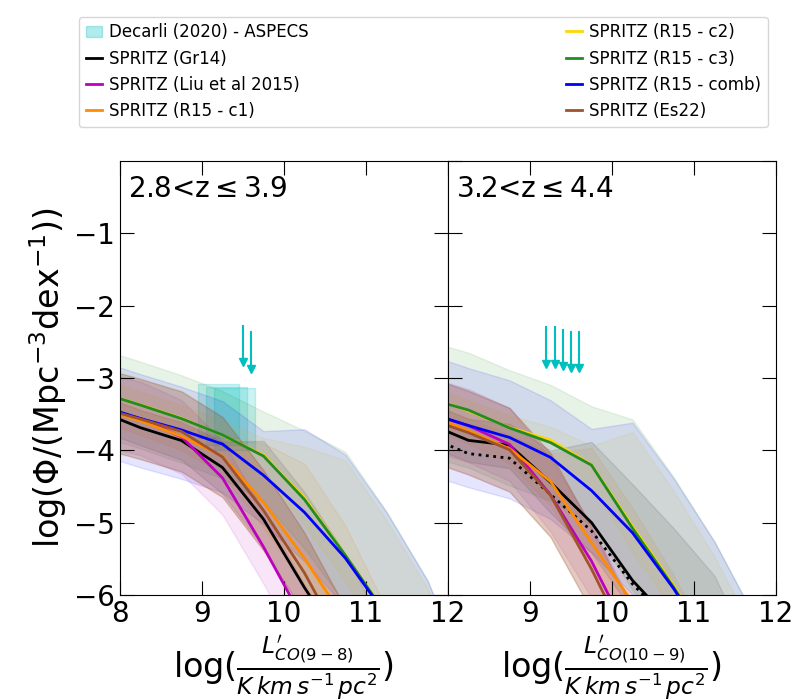}
    \caption{The same as Figure \ref{fig:LCO_1}, but for CO(4-3) (\textit{top left}), CO(5-4) and CO(6-5) (\textit{top right}), CO(7-6) and CO(8-7) (\textit{bottom left}), CO(9-8) and CO(10-9) (\textit{bottom right}). R15 refers to the relations by \citet{Rosenberg2015}}.
    \label{fig:LCO_4_10}
\end{figure*}

In Figure \ref{fig:LCO_4_10} we report the comparison between the results from \spr{} and the observed CO LFs from ${J}=4$ to ${J}=10$. We tested different relations inside \spr{} to estimate the mid-$J$ CO LFs, as described in Sec. \ref{sec:CO_relations} and summarised in Table \ref{tab:CO_references}. \par
The faint-end slopes of the CO(4-3) LFs are similar for all the considered relations and they are within 0.5 dex even for ${J}>4$. The main differences between the considered predictions are, therefore, on the knee position and the bright-end slope of the LFs. The CO(4-3) observational data cover a narrow luminosity range around the knee at $z\sim1$, with all the considered relations slightly below the observations, but still consistent within the error bars. This may be linked to the light offset between the input \spr{} IR LF and the observed one at $z=1.0-1.7$ (see Appendix \ref{sec:LIR}). At $z=3.5$--3.9 the \spr{} and the observed LFs have different shapes, with the knee of the latter being at lower luminosities and at higher densities, and the bright-end slope being steeper than the ones predicted by \spr{} with any relation. This discrepancy has no obvious explanation and more observations over a larger luminosity range (ASPECS observations correspond to a single independent luminosity bin) and at additional redshfits are needed to investigate it further. \par
For the CO(5-4) and CO(6-5) the observed LFs are generally higher than the different relations included in \spr. One exception is the LF derived using the relation by \citet{Liu2021}, which includes a dependence of the CO(5-4)/CO(2-1) ratio on the IR luminosity. This may indicates that the different CO transitions may have different dependence with the IR luminosity. However, the observed underestimation of the CO(5-4) LF may be linked to the observed underestimation of the CO(1-0) LF at the same redshift (i.e $z=1.2-1.6$, see previous Section and Appendix \ref{sec:LIR}). No CO(1-0) observations are instead available at $z=1.7-2.1$ to investigate further the offset observed for the CO(6-5), but we highlight that the observed IR LF at these redshift are perfectly reproduced in \spr, therefore we would expect to equally reproduce the CO(1-0) (assuming the CO(1-0)-L$_{IR}$ does not strongly evolve with redshift). Finally, for transitions with ${J}>6$ all the predictions are consistent with the observations within the error bars. \par
For comparison, the models by \citet{Popping2016} generally follow the observed LFs for ${J}=4$ to 6, with some discrepancies at $z>2$, which is present only for ${J}=4$. As mentioned in the previous section, the same model strongly differs also from the observed LFs for ${J}<3$ at the same redshifts.\par
Overall, looking at all the high-$J$ LFs and the available observations, the best agreement for \spr{} is obtained by considering the CO($J\rightarrow(J-1)$)/\cii ratios for the different classes presented in \citet{Rosenberg2015}, as they generally result in smaller $\chi^{2}$ values than the other relations (see Table \ref{tab:chi2_CO}). For the classes by \citet{Rosenberg2015}, with increasing redshift or J-value, the inclusion of only high-excitation classes (c2 and c3) seems to be preferred by the observations. However, more observations, for instance that of ${J}>7$ at $z<1$, are necessary to discriminate whether the better agreement with the high-excitation classes is driven by the redshift evolution (i.e. the excitation increases with redshift) or by the $J$-value (i.e. the excitation increases with $J$-value) or a combination of both. At the same time, more observations of the bright-end (L$^{\prime}_{CO(4-3)}>10^{10} {\rm K\,km\,s}^{-1}$pc$^{2}$) are necessary to further discriminate among the different relations considered.
\par
 
\section{Molecular gas mass}\label{sec:H2}
As previously mentioned, CO can be efficiently used to trace the molecular content of a galaxy. However, sometimes $\rm H_2$ may survive outside the CO regions, in the so-called CO-dark clouds \citep[e.g.][]{Wolfire2010}, and can be instead efficiently traced by the [CII] emission \citep[e.g.][]{Zanella2018,Madden2020, Wolfire2022}. For this reason, to calculate the molecular gas mass of each simulated galaxy in \spr, we decided to use both proxies. \par

\begin{figure*}
    \centering
    \includegraphics[width=0.8\linewidth,keepaspectratio]{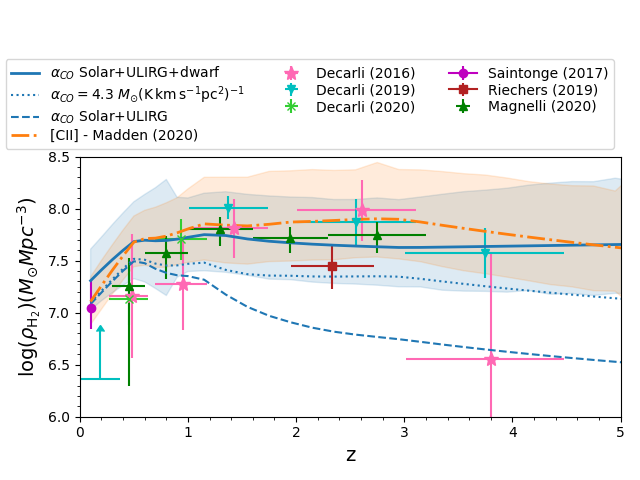}
    \caption{Molecular gas mass density derived by \spr{} from the \cii, using the \cii luminosity derived from \citetalias{Vallini2015} and the relation by \citet[][\textit{orange dash-dotted line}]{Madden2020}. We also report the $\rm H_2$ mass density derived from the CO, using the CO(1-0) luminosity derived from \citet{Greve2014} and \citet{Sargent2014} and considering different $\alpha_{CO}$ values for star-forming galaxies, ULIRG-like and dwarfs (\textit{blue solid line}). We also show the molecular gas mass density obtained by considering a single $\alpha_{CO}$ value for all the galaxies (\textit{blue dotted line}), or two different $\alpha_{CO}$ values for star-forming and ULIRG-like galaxies (\textit{blue dashed line}). We report observational results by \citet{Decarli2016,Decarli2019,Decarli2020}, \citet{Saintonge2017}, \citet{Riechers2019}, and \citet[][from the evolution of the dust mass density]{Magnelli2020}. The shaded areas include the 1$\sigma$ uncertainties on the input LF, the high-z extrapolations and the $\rm H_2$ conversions. For the sake of clarity, we report uncertainties only for the results obtained from \cii and from CO, assuming three different $\alpha_{CO}$ values for star-forming galaxies, ULIRG-like and dwarfs. The uncertainties associated to other results derived from CO, under different $\alpha_{CO}$ assumption, are comparable.}
    \label{fig:MH2}
\end{figure*}

In particular, on the one hand we derived the molecular gas mass directly from the CO(1-0) by considering the Milky Way value $\alpha_{CO}=4.3$ $\rm M_{\odot}({\rm K\,km\,s}^{-1}{\rm pc}^{2})^{-1}$ for normal star-forming galaxies (Spiral and SF-AGN) and the value derived for ULIRGs ($\alpha_{CO}=0.86\,\rm M_{\odot}({\rm K\,km\,s}^{-1}{\rm pc}^{2})^{-1}$) for the most active galaxies (intense star-formation or nuclear activity, i.e. SB, SB-AGN, AGN1 and AGN2). For dwarf galaxies we considered a metallicity-dependent CO-to-$\rm H_2$ conversion factor, as derived by \citet{Madden2020}, i.e. $\alpha_{CO}=10^{0.58}\times(Z/Z_{\odot})^{-3.39}$, by taking into account the contribution of the CO-dark clouds. We considered as reference the CO(1-0) luminosity estimated from the IR emission using the relation by \citet{Sargent2014} for star-forming galaxies, with the correction by \citet{Hunt2015} for galaxies with sub-solar metallicity, and by \citet{Greve2014} for AGN-dominated systems (see Sec. \ref{sec:CO_relations}). \par
On the other hand, we calculated the molecular gas mass from the \cii line luminosity as M$_{\rm H_2}=10^{2.12}(L_{[\ion{C}{II}]}/{\rm L}_{\odot})^{0.97}$ \citep{Madden2020}, without any variation among different galaxy populations. In this case we considered the \cii luminosity estimated using the relation by \citet{Vallini2019}, which is the one with the best agreement with the observations (see Sec. \ref{sec:cii_obs}).

\subsection{The molecular gas mass density: comparison with observations}

To validate the molecular gas mass included in \spr, derived either from CO or from \cii for each simulated galaxy, we estimate the cosmic evolution of the molecular gas mass and compare it with available observations (Figure \ref{fig:MH2}). \par
On the one hand, when comparing the molecular gas mass density derived from the predicted CO(1-0) luminosity, it is evident that using the Milky Way $\alpha_{CO}$ value for all the galaxies leads to underpredict by $\sim$0.5 dex the molecular gas mass at $z>0.5$. This underestimation increases up to 1 dex, if we assign a lower $\alpha_{CO}$ value to the most active galaxies (intense star-formation or nuclear activity, i.e. SB, SB-AGN, AGN and AGN2), but it is balanced if we include an $\alpha_{CO}$ that varies with metallicity in dwarf galaxies. Therefore, the best option to convert the CO into molecular gas mass seems to be represented by the use of a different $\alpha_{CO}$ for normal star-forming galaxies, active systems and dwarf galaxies. \par
On the other hand, a single $\alpha_{\ion{C}{II}}$ value seems to be enough to reproduce the observed shape of the molecular gas mass density, showing a peak around $z=2$, then decreasing at lower and higher redshfit. A value of $\alpha_{\ion{C}{II}}\sim$130, as proposed by \citet{Madden2020}, is necessary to reproduce the observed normalisation with the \spr{} simulation.\par
\begin{figure}
    \centering
    \includegraphics[width=0.99\linewidth,keepaspectratio]{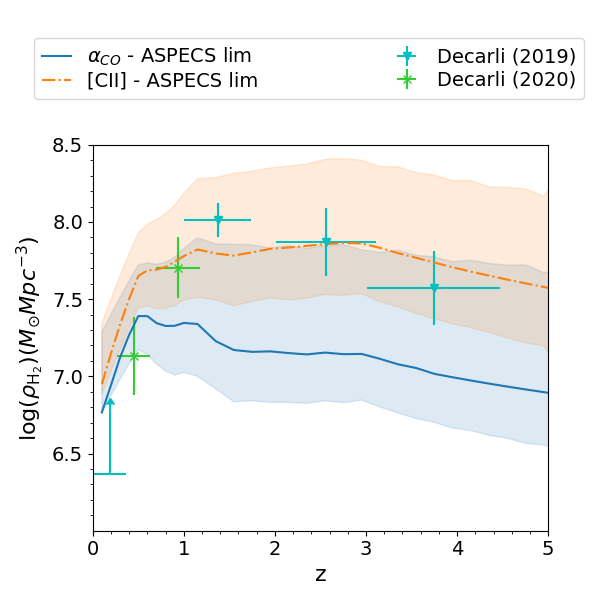}
    \caption{Molecular gas mass density derived by \spr{} by considering the ASPECS observational limits. The results are derived from both the \cii (using the relation by \citealt[][\textit{orange dash-dotted line}]{Madden2020}), and from the CO (considering $\alpha_{CO}$=3.6 {\rm}M$_{\odot}$(K kms$^{-1}$ pc$^{2})^{-1}$, as in the ASPECS survey; \textit{blue solid line}).}
    \label{fig:MH2_Alim}
\end{figure}

For a proper comparison with observations is necessary to apply the observational limits of the ASPECS survey\footnote{We also considered the ASPECS $\alpha_{CO}=3.6\,{\rm M}_{\odot}({\rm K\,km\,s}^{-1}{\rm pc}^{-2})^{-1}$, but a small change in the $\alpha_{CO}$ value has a minor impact on the results.}, as their molecular mass density is derived \textit{without} extrapolating their CO LF.
For example, once the ASPECS observational limits have been applied, the model by \citet{Popping2019} is a factor of 2-3 lower than the observations \citep{Popping2019b}. Similar or even larger discrepancies are present in \spr{} when deriving the molecular gas from the CO, as seen in Figure \ref{fig:MH2_Alim}. This discrepancy is present because the majority of dwarf galaxies are below the ASPECS observational limits. Taking into account that the ASPECS $\rm H_2$ values were derived from the CO(1-0) to CO(4-3) line luminosities, the discrepancies present in the $\rm H_2$ may arise from the differences observed for the $J=2$ and $J=4$ transitions (see Section \ref{sec:COobs}). \par
The molecular gas mass density derived from \cii is instead well consistent with the observations, even after applying the observational limit of the ASPECS survey. Observational data point out to a light decrease of the molecular gas mass at $z>2$, while in \spr{} the contribution of SB and SB-AGN (see Figure \ref{fig:LCII_SED_z}) keep the molecular gas density almost constant. However, the big uncertainties do not allow for investigating this further. Therefore, we can conclude that with \spr{} it is possible to estimate a reliable molecular gas mass density starting from the \cii, at least up to $z=4$.

\section{Summary and conclusions}\label{sec:conclusions}
In this work we used the state-of-the-art \spr{} simulation to predict the \cii and CO LFs at different redshifts as well as the molecular gas mass density. In particular, we have considered different relations from the literature to derive the \cii luminosities \citep{DeLooze2014,Vallini2015,Gruppioni2016}, starting from the IR luminosity or the SFR. For the CO transitions with $J<14$, we tested different relations, starting from the IR luminosity or based on the CO($J\rightarrow(J-1)$)/CO(1-0) or CO($J\rightarrow(J-1)$)/\cii ratios, in particular those by \citet{Sargent2014}, \citet{Greve2014}, \citet{Rosenberg2015}, \citet{Liu2015,Liu2021}, \citet{Boogaard2020} and \citet{Esposito2022}. In addition, we included predictions for the LFs of CO with $J=14$--24 using the CO SLED of galaxies observed by \citet{Mashian2015} as templates. However, no observed LFs are currently available to test the predictions for CO transitions with $J\geq14$. \par
For the \cii, at $z<0.3$ the best result is obtained considering the relation by \citetalias{Vallini2015}, which takes into account not only a dependence of \cii on SFR, but also on metallicity. The relation is also consistent with observations at high-\textit{z}, where, however, more observations at ${\rm L}_{[\ion{\rm C}{II}]}>10^{10}{\rm L}_{\odot}$ are necessary to unambiguously discriminate between the considered relations. Future IR spectroscopic observations (from space, given the Earth atmospheric transmission), covering wavelengths shorter than those sampled by ALMA, will be essential to explore intermediate redshifts and provide valuable constraints for the different relations. \par
For low-$J$ values ($J\leq3$), the CO LFs in \spr, derived using the relations by \citet{Sargent2014}, \citet{Greve2014}, \citet{Boogaard2020} or \citet{Esposito2022} for different galaxy populations, are in good agreement with the observations. The only discrepancy occurs for the CO(2-1) transition at $z=1.2$--1.6, where the \spr{} LF is slightly below the observed data, but still consistent within the errors. \par
For mid- and high-$J$ CO transitions (i.e., $J>3$), the best results are obtained by considering the CO/\cii ratios derived for the different classes by \citet{Rosenberg2015}, while for the CO(5-4) LF, the relation by \citet{Liu2021}, including a further dependence of the CO(5-4)/CO(2-1) ratio on IR luminosity, provides one of the best results, when compared with the available observations. However, all relations are generally consistent with each other in the faint-end and more observations at luminosities brighter than $\sim10^{10.5} {\rm K\,km\,s}^{-1}{\rm pc}^{2}$ are necessary to unambiguously discriminate between the predictions of the different models. \par
Finally, we integrate the CO and \cii LFs, after converting them to molecular gas mass through different recipes, to obtain an estimate of the molecular gas mass density at different redshifts. The evolution of the molecular gas mass density is correctly reproduced by \spr{} over the whole redshift range where observations are available (i.e., $0<z<4$), in particular by deriving the $\rm H_2$ mass directly from the \cii LF. \par
We conclude that the \spr{} simulation can be used to predict the evolution of both the \cii and CO luminosity, as well as that of the molecular gas mass. This work constitutes a useful reference for any future sub-mm/mm observations and strongly outlines the need of a future far-IR spectroscopic instrument that covers the huge gap between the past \hers{} observations ($\lambda\leq210\,\mu$m) and the current ones with ALMA ($\lambda\geq300\,\mu$m). This would be fundamental to obtain statistical samples of galaxies over a continuous redshift range and derive a better understanding of the quantities discussed in this paper.
 

\begin{acknowledgements}
We thanks M. B\'{e}thermin for the useful comments. LB, CG acknowledge the support from grant PRIN MIUR 2017-20173ML3WW$\_$001. LB acknowledges also support from grant PRIN2017. This work has been partially funded by Premiale MITIC 2017. LV acknowledges support from the ERC Advanced Grant INTERSTELLAR H2020/740120 (PI: Ferrara).
\end{acknowledgements}

%
\bibliographystyle{aa} 
\bibliography{main} 
%

\begin{appendix} 
\section{\spr{} updates}\label{sec:newspritz}
We update the redshift evolution of the GSMF of dwarf irregular galaxies in \spr, particularly at $z \lesssim 0.5$, by complementing the values derived by \citet{Huertas-Company2016}, which were included in the original \spr{} \citepalias[v1;][]{Bisigello2021}, with the local values derived by \citet{Moffett2016}.  In both works, the GSMF has been described by a single Schechter function and the same functional form has been kept in \spr. \par
\begin{figure}
    \centering
    \includegraphics[width=0.98\linewidth,keepaspectratio]{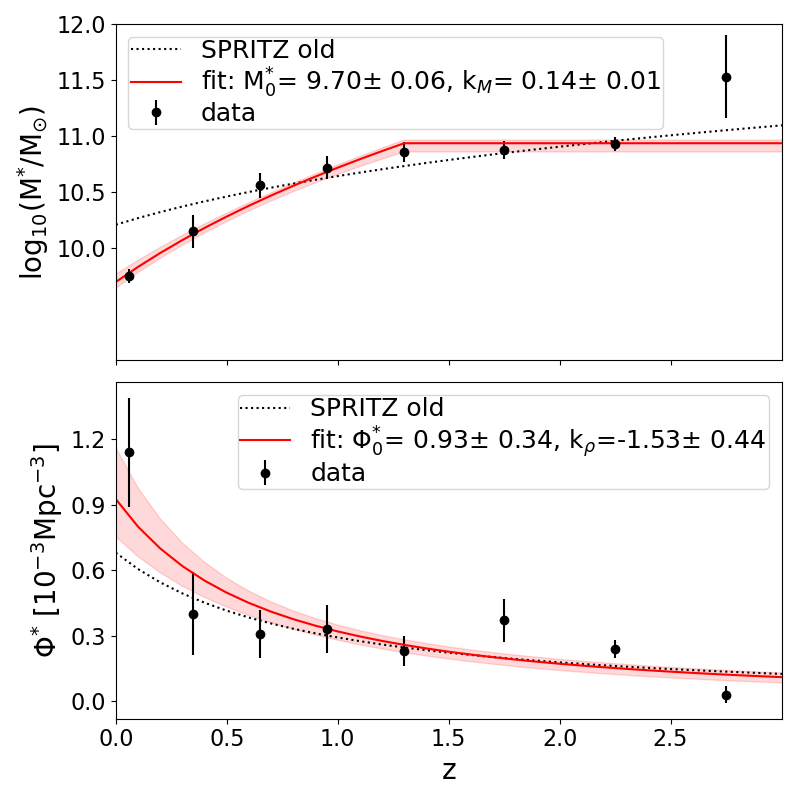}
    \caption{Redshift evolution of the characteristic stellar mass (\textit{top}) and the characteristic density (\textit{bottom}) of the GSMF of dwarf irregular galaxies. Black circles are the observed values by \citet[][$z>0.06$]{Huertas-Company2016} and \citet[][$z\sim0.06$]{Moffett2016}. The red lines and shaded areas show the best fit and the corresponding uncertainties, while the black dotted lines indicate the evolution of the GSMF previously present in \spr. }
    \label{fig:GSMF_Irr}
\end{figure}

To describe the redshift evolution of this galaxy population, the number density at the knee $\Phi^{*}$ and the characteristic stellar mass $M^{*}$ at $z\leq1.3$ are both considered as functions of redshift, evolving as $\propto(1+z)^{k}$. The best-fitting results are shown in Figure \ref{fig:GSMF_Irr} and they correspond to $\log(M^{*}/M_{\odot})=(9.70\pm0.06)\,(1+z)^{0.14\pm0.01}$ at $z\leq1.3$ and $\Phi^{*} [10^{-3} Mpc^{-3}]=(0.93\pm0.34)\,(1+z)^{-1.53\pm0.44}$. The characteristic stellar mass is kept constant at $z>1.3$, as no particular redshift evolution is observed above this redshift. \par
\begin{figure}
    \centering
    \includegraphics[width=0.98\linewidth,keepaspectratio]{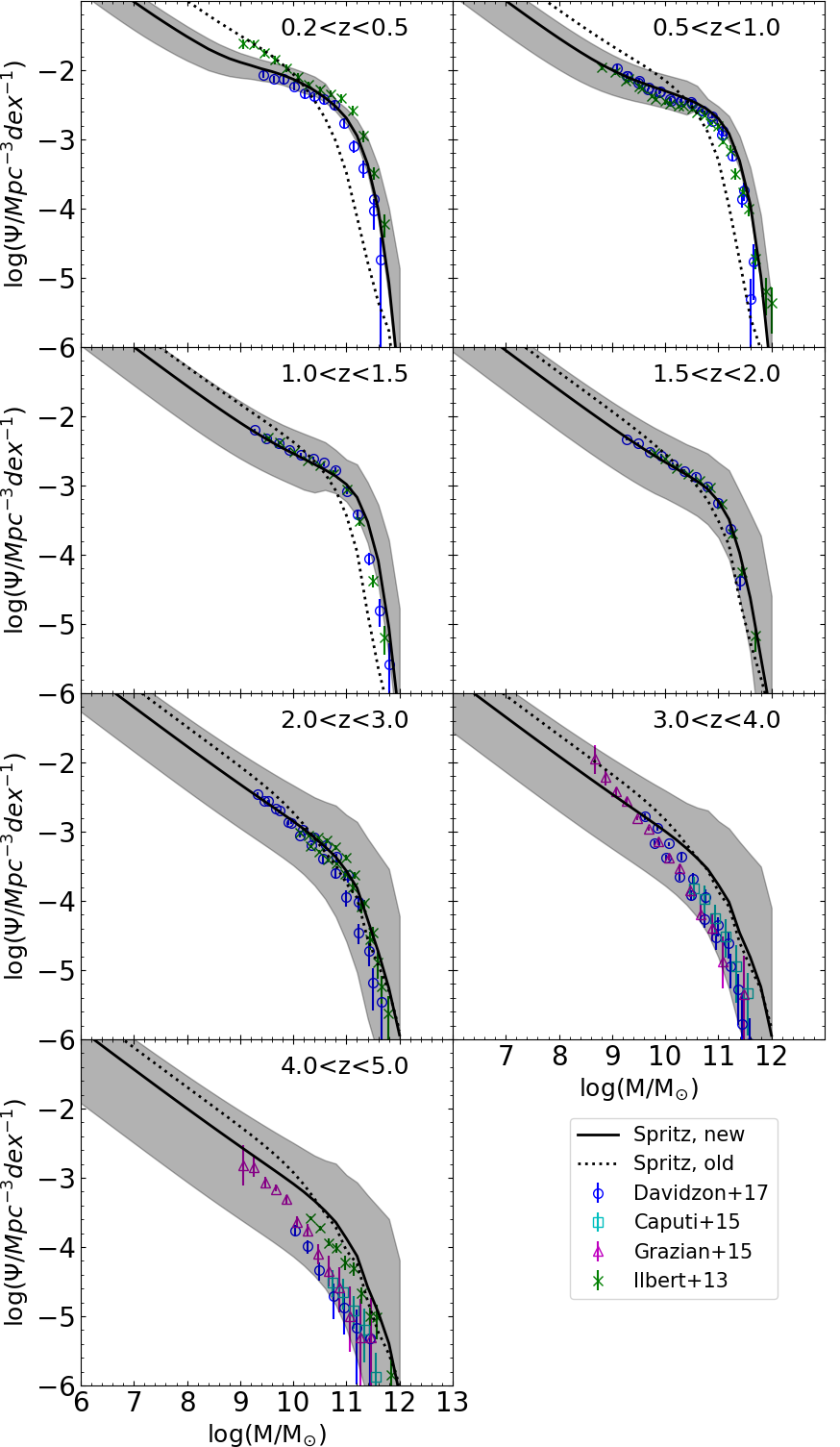}
    \caption{\spr{} total GSMF at different redshifts before (\textit{dotted black line}) and after (\textit{thick solid black line}) applying the changes on the GSMF of dwarf irregular galaxies. We also report observational values present in the literature: \citet[\textit{green crosses}]{Ilbert2013}, \citet[\textit{cyan squares}]{Caputi2015}, \citet[][\textit{magenta triangles}]{Grazian2015}, and \citet[\textit{blue circles}]{Davidzon2017}.}
    \label{fig:newMF}
\end{figure}

The proposed change produces a decrease on the reduced $\chi^{2}_{red}$ of the characteristic stellar mass $M^{*}$ from 70.3 to 5.7, and on the reduced $\chi^{2}_{red}$ of the number density at the knee $\Phi^{*}$ from 18.4 to 16.9. In addition, the low-mass end of the total GSMF in \spr{} is now in much better agreement with the available observations at all redshifts, as can be appreciated in Figure \ref{fig:newMF}.

\section{Propagation of the UV LF discrepancies on the \cii LF}\label{sec:SFRUV}
\begin{figure}
    \centering
    \includegraphics[width=\linewidth,keepaspectratio]{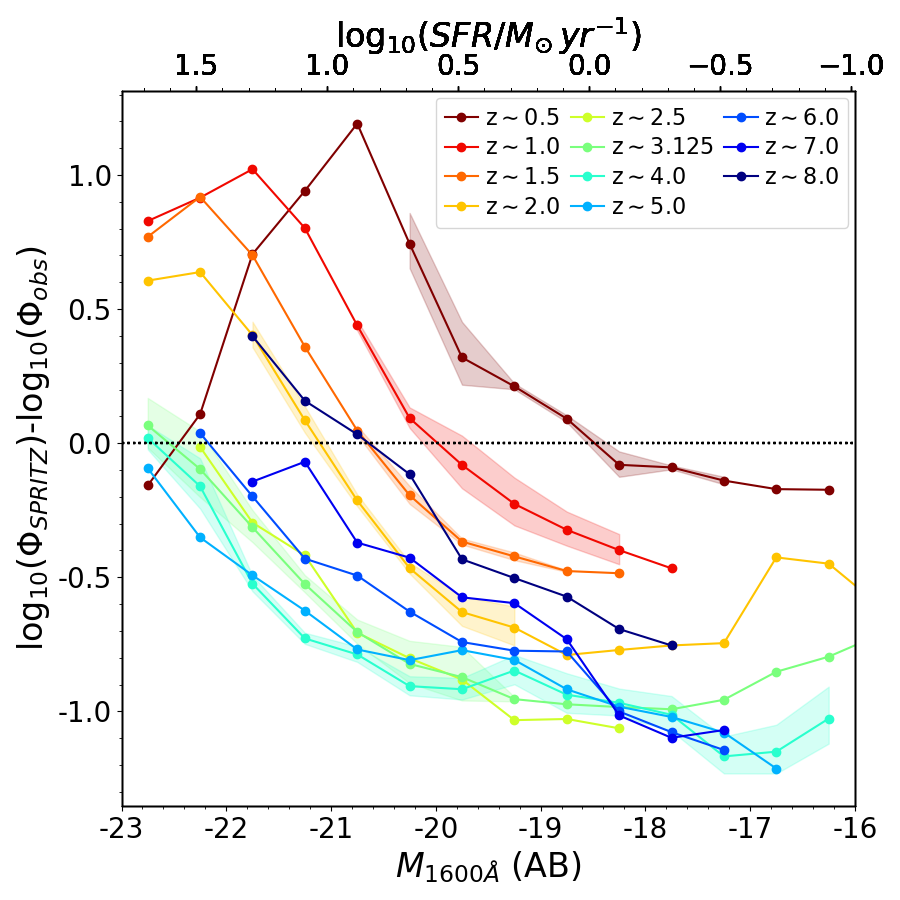}
    \caption{Difference between the \spr{} and the average observed UV LFs as a function of UV magnitude (bottom axis) or UV SFR (top axis). Shaded areas show the 1$\sigma$ scatter of the observed UV LFs, when multiple estimates are available. Colors indicate different redshifts.}
    \label{fig:DLFUV}
\end{figure}

In this section we investigate the impact of the discrepancies between the simulated and the observed UV LF on the derived \cii LF. We do not attempt to derive a precise correction to apply to the \cii luminosity of each simulated galaxy in \spr, but instead we estimate an average correction to the \cii LF to have a indication of possible biases. \par
In Figure \ref{fig:DLFUV} we report the difference between the \spr{} and the observed UV LF. We also included the difference with respect to the UV component of the SFR function, derived using the \citet{Kennicutt1998a} relation. As a simple approach, we consider as reference observed LF the average among the results by \citet{Reddy2009}, \citet{Oesch2010}, \citet{Bouwens2015}, \citet{Parsa2016}, \citet{Adams2020} and \citet{Moutard2020},  but in the same Figure we also report the scatter among the different observed values. We consider only M$_{1600\text{\AA}}<-23$ as the LF is dominated at brighter magnitudes by un-obscured AGN, whose UV emission is predominately originated by the accretion disk of the AGN and not by SFR. As mentioned in \citetalias{Bisigello2021}, the bright-end of the galaxy UV LF at $z<2$ is overestimated, mainly because of an excess of spiral galaxy. On the other hand, the faint-end is generally underestimated, probably because of the limited amount of SED used in the simulation or to an additional dust-poor population not observed by \hers or included in the dwarf irregulars. The difference in the faint-end increases with redshift. \par

\begin{figure*}
    \centering
    \includegraphics[width=0.78\linewidth,keepaspectratio]{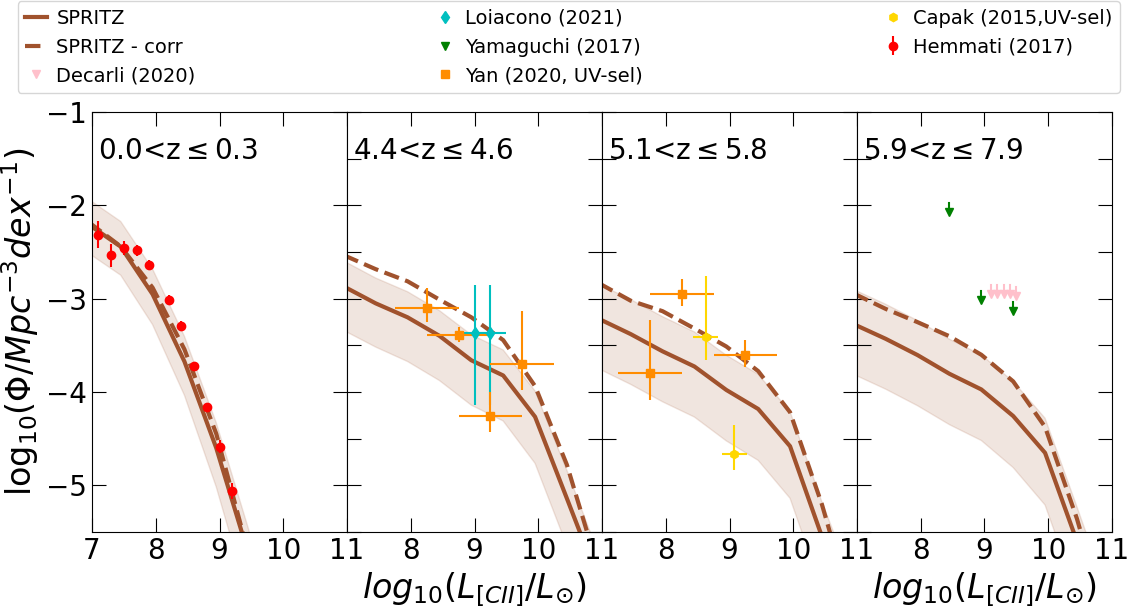}
    \caption{The same as Figure \ref{fig:LCII}, but with the \cii obtained considering the relation by \citetalias{Vallini2015} before (\textit{continuous lines}) and after (\textit{dashed lines}) the correction estimated from the UV LF. }
    \label{fig:CII_corr}
\end{figure*}

Then, we derive the median SFR$_{UV}$ corresponding to different IR luminosities in the simulation and we use it to apply the difference in the M$_{1600\text{\AA}}$ LF previously derived to the \cii LF. In Figure \ref{fig:CII_corr} we report the un-corrected and the corrected \cii LF derived considering, for example, the relation by \citetalias{Vallini2015}. The main difference is at high-$z$, with an increase of even 0.4 dex in the faint-end. This correction is at the edge of the derived uncertainties for the \cii LF and overestimated some observations. Overall, this simple test is useful to verify that the discrepancies between the \cii and CO LFs may at least be due to an underestimation of the SFR$_{UV}$ in \spr.

\section{CO SLEDs}\label{sec:co_sled}
In this section we show the CO SLED associated to each galaxy type using the different relations reported in Table \ref{tab:CO_references}. The CO SLED of AGN2 is identical to the one of AGN1, as they have been derived from the same IR LF \citepalias[see][]{Bisigello2021} and we use the same relations to derive the different CO luminosities (see Table \ref{tab:CO_references}).\par
The CO SLED of spiral galaxies is generally smooth, with the different relations giving values within a factor of three for the same J values. The same happens for the SB CO SLED at J$\leq$13. However, the L$_{CO(14-13)}^{\prime}$/L$_{CO(1-0)}^{\prime}$ derived from the work by \citet{Mashian2015} is a factor of 10 larger than the L$_{CO(13-12)}^{\prime}$/L$_{CO(1-0)}^{\prime}$ derived from \citet{Greve2014}, producing a discontinuity on the overall CO SLED. Unfortunately, in the work by \citet{Mashian2015} there are no CO SLEDs steeper than the one of M82, which is used as a template for SB galaxies, therefore it is not possible at the moment to improve the agreement with the CO SLED by \citet{Greve2014}. \par
The CO SLED of dwarf irregular galaxies is generally smooth when considering the relations by \citet{Greve2014,Liu2015} and \citet{Boogaard2020}, while the CO SLED by \citet{Rosenberg2015} is 5-10 times higher than the others. This is at least partially due to the different metallicity dependence considered when deriving the \cii and CO luminosity. In general, further observations of the CO SLED of dwarf galaxies are necessary to disentangle between the different models. \par 
Moving to galaxies with an AGN component, the CO SLED of SF-AGN follows two trends as the CO SLEDs by \citet{Boogaard2020}, \citet{Greve2014} and \citet{Liu2015} are between 5 and 20 times lower than the CO SLED by \citet{Esposito2022}, which is consistent with the one by \citet{Rosenberg2015}. We remind the reader that the relations by \citet{Esposito2022} and \citet{Rosenberg2015} are specific for objects hosting a low-luminosity AGN and should be, therefore, more suited to describe SF-AGN, while the other relations are broadly valid for star-forming galaxies. Moreover, for galaxies at high redshift (i.e., $z>1.5$), which should be more contaminated by AGN, the L$_{CO(J-(J-1))}^{\prime}$/L$_{CO(1-0)}^{\prime}$ ratio by \citet{Boogaard2020} increases becoming closer to the ratios by \citet{Esposito2022}.  \par
Finally, AGN1, AGN2 and SB-AGN show similar CO SLEDs with a large scatter (i.e. up to a factor of 13) between the different relations. For $J>14$, the CO SLED by \citet{Mashian2015} is completely in agreement with the CO SLED by \citet{Esposito2022} for AGN1 and AGN2, while it is a factor of five lower than the same CO SLED for SB-AGN. The CO SLED derived using results from \citet{Greve2014} is always the steeper one, showing the smallest L$_{CO(J-(J-1))}^{\prime}$/L$_{CO(1-0)}^{\prime}$ values.\par
Overall, switching from one relation to another to estimate the CO luminosity at different J values may produce, in some cases, large discontinuities on the estimated CO SLED. In the future, observations of statistical samples of galaxies over wide J ranges may be used to improve the CO SLED included in \spr{}.

\begin{figure*}
    \centering
    \includegraphics[width=0.48\linewidth,keepaspectratio]{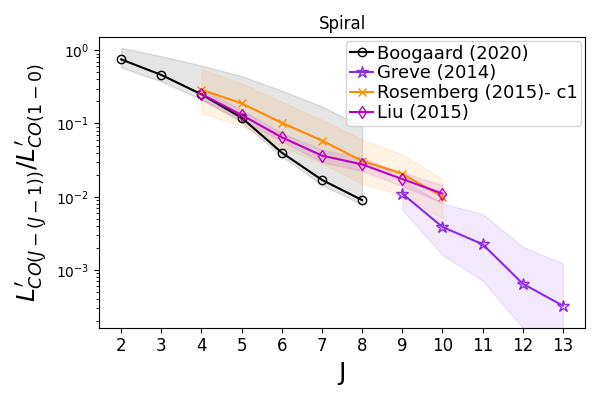}
    \includegraphics[width=0.48\linewidth,keepaspectratio]{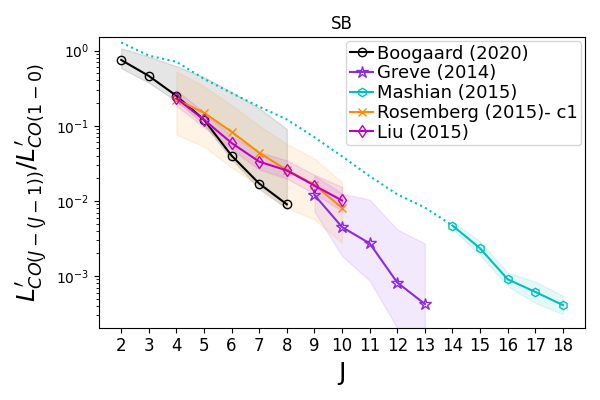}
    \includegraphics[width=0.48\linewidth,keepaspectratio]{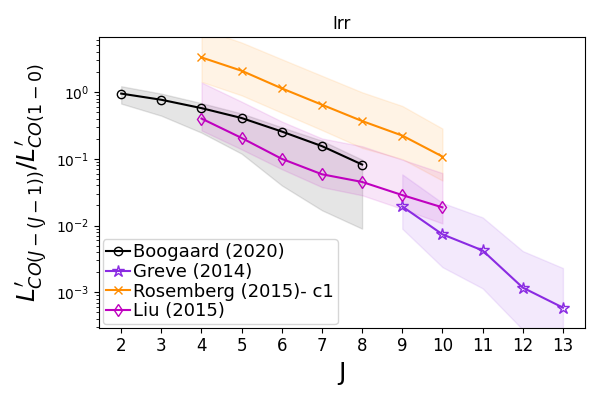}
    \includegraphics[width=0.48\linewidth,keepaspectratio]{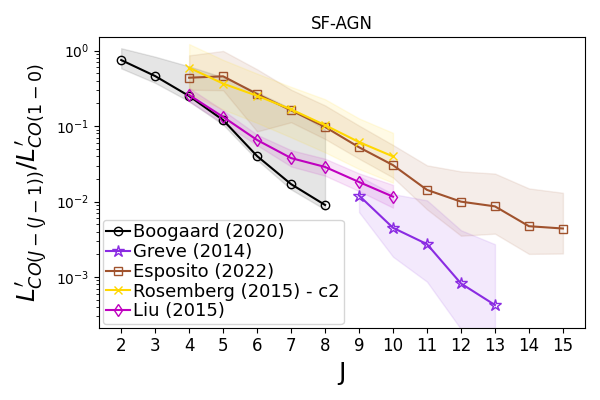}
    \includegraphics[width=0.48\linewidth,keepaspectratio]{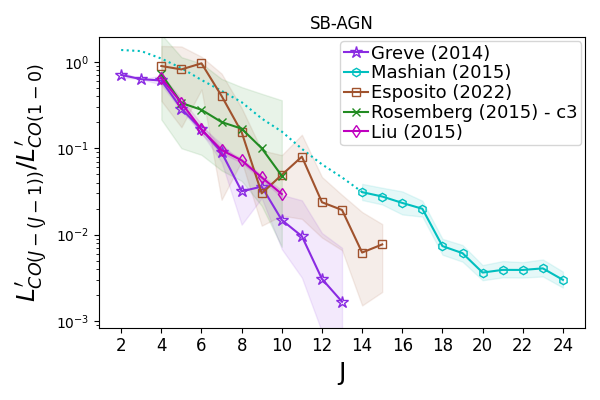}
    \includegraphics[width=0.48\linewidth,keepaspectratio]{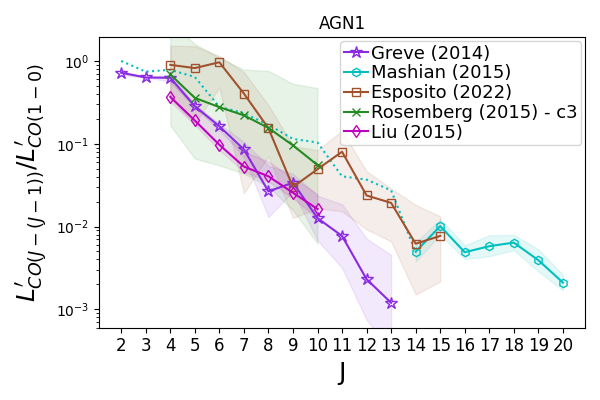}
    \caption{Median CO SLED of the different galaxy types included in \spr{}: spiral, SB, Irr, SF-AGN, SB-AGN and AGN1. The CO SLED of AGN2 is identical to the one of AGN1 and is not shown. Different symbols indicate the relations considered to derive the CO luminosity for different J. The dashed cyan dotted line shows the CO SLED by \citet{Mashian2015} for $J<14$, which are not included in this work but are shown for consistency. Shaded areas show the one $\sigma$ variation of the CO SLED for galaxies with $10^{10}\leq L_{IR}/L_{\odot}\leq10^{12}$.}
    \label{fig:COSLED}
\end{figure*}

\section{\cii LF up to $z=10$}\label{sec:cii_allz}
In this section we report, as reference for future IR observations, the \cii LFs derived in \spr. \par
\begin{figure}
\centering
    \includegraphics[width=0.75\linewidth,keepaspectratio]{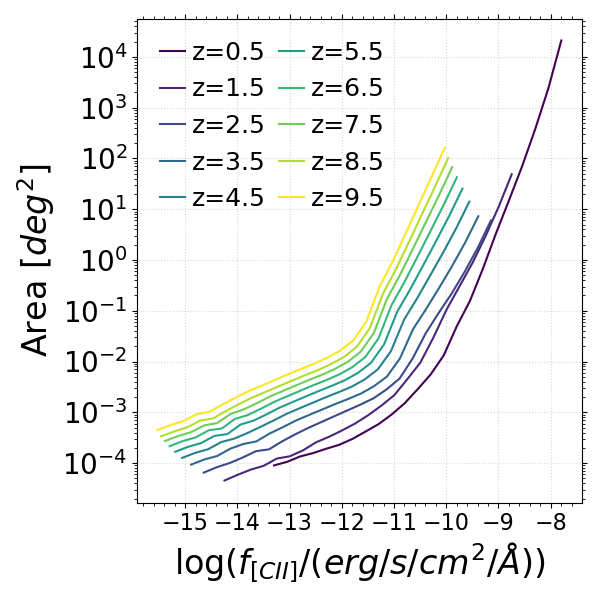}
    \caption{Flux corresponding to \cii luminosity between 10$^{7}$ and 10$^{11}$ L$_{\odot}$, between $z=0.5$ and $z=9.5$, compared with the area necessary to observe at least one object.}
    \label{fig:Area_depth_cii}
\end{figure}

To derive the \cii luminosity of each simulated galaxy we considered the relation by \citetalias{Vallini2015}, given the agreement with the results obtained with this relation and the available observations (see Sec. \ref{sec:cii_obs}). We report in Table \ref{tab:LCII_zall} the derived values of the \cii LFs up to $z=10$. As reference for future far-IR spectroscopic missions, we also report in Figure \ref{fig:Area_depth_cii} the line flux corresponding to different \cii luminosities, ranging from 10$^{7}$ to 10$^{11}$ L$_{\odot}$, and the area necessary to observe at least one object, looking at the predicted \cii LF. We report line fluxes, instead of line luminosities as done in the rest of the manuscript, to allow for a direct and easy comparison with future instrument performances. For example, to detect the \cii line of an object in the knee of the \cii LF at $z\sim2.5$, it will be necessary to observe down to 10$^{-11}$ erg$\,$s$^{-1}\,$cm$^{-2}\,$\AA$^{-1}$ and an area larger than $\sim$4$\times10^{-3}$ deg$^{2}$. This estimate takes into account only the flux of the obeject and it needs to be tuned to the desired S/N value. \par

\begin{table*}
    \caption{\cii LF predicted by \spr{} between $<z>=0.5$ and $<z>=9.5$, considering a luminosity bin of 0.25 dex. Errors (1$\sigma$) include the uncertainties on the input LFs or GSMF and the scatter around the relation.  }
    \centering
    {\renewcommand{\arraystretch}{1.5}%
    \resizebox{0.99\textwidth}{!}{
    \begin{tabular}{c|cccccccccc}
    $<{\rm log}(L_{[\ion{\rm C}{II}]}/L_{\odot})>$ & \multicolumn{10}{c}{${\rm log}(\Phi/Mpc^{-3}dex^{-1})$} \\
 & $<z>=0.5$  & $<z>=1.5$ & $<z>=2.5$ & $<z>=3.5$ & $<z>=4.5$ & $<z>=5.5$ & $<z>=6.5$ & $<z>=7.5$ & $<z>=8.5$ & $<z>=9.5$ \\
 \hline 
7.125 & -2.290$^{+0.270}_{-0.364}$ & -2.559$^{+0.272}_{-0.437}$ & -2.827$^{+0.272}_{-0.459}$ & -2.996$^{+0.274}_{-0.469}$ & -3.125$^{+0.273}_{-0.472}$ & -3.230$^{+0.272}_{-0.475}$ & -3.270$^{+0.372}_{-0.538}$ & -3.421$^{+0.373}_{-0.539}$ & -3.432$^{+0.373}_{-0.538}$ & -3.611$^{+0.371}_{-0.541}$\\ 
7.375 & -2.371$^{+0.270}_{-0.361}$ & -2.663$^{+0.273}_{-0.436}$ & -2.918$^{+0.273}_{-0.457}$ & -3.089$^{+0.273}_{-0.471}$ & -3.177$^{+0.273}_{-0.473}$ & -3.327$^{+0.274}_{-0.475}$ & -3.413$^{+0.373}_{-0.540}$ & -3.422$^{+0.373}_{-0.539}$ & -3.566$^{+0.373}_{-0.540}$ & -3.629$^{+0.372}_{-0.540}$\\ 
7.625 & -2.464$^{+0.270}_{-0.361}$ & -2.761$^{+0.272}_{-0.438}$ & -2.981$^{+0.275}_{-0.459}$ & -3.180$^{+0.272}_{-0.472}$ & -3.293$^{+0.274}_{-0.475}$ & -3.378$^{+0.272}_{-0.477}$ & -3.491$^{+0.373}_{-0.539}$ & -3.586$^{+0.373}_{-0.542}$ & -3.589$^{+0.371}_{-0.541}$ & -3.718$^{+0.372}_{-0.541}$\\ 
7.875 & -2.586$^{+0.270}_{-0.366}$ & -2.855$^{+0.271}_{-0.440}$ & -3.088$^{+0.274}_{-0.462}$ & -3.267$^{+0.273}_{-0.472}$ & -3.371$^{+0.274}_{-0.476}$ & -3.473$^{+0.273}_{-0.478}$ & -3.565$^{+0.372}_{-0.541}$ & -3.656$^{+0.374}_{-0.541}$ & -3.735$^{+0.372}_{-0.542}$ & -3.806$^{+0.372}_{-0.543}$\\ 
8.125 & -2.734$^{+0.270}_{-0.373}$ & -2.950$^{+0.272}_{-0.442}$ & -3.217$^{+0.272}_{-0.466}$ & -3.327$^{+0.274}_{-0.472}$ & -3.504$^{+0.273}_{-0.479}$ & -3.551$^{+0.274}_{-0.479}$ & -3.648$^{+0.372}_{-0.542}$ & -3.740$^{+0.373}_{-0.543}$ & -3.828$^{+0.374}_{-0.542}$ & -3.911$^{+0.372}_{-0.543}$\\ 
8.375 & -2.918$^{+0.270}_{-0.386}$ & -3.067$^{+0.273}_{-0.446}$ & -3.328$^{+0.272}_{-0.466}$ & -3.427$^{+0.272}_{-0.474}$ & -3.577$^{+0.274}_{-0.478}$ & -3.690$^{+0.273}_{-0.482}$ & -3.728$^{+0.374}_{-0.543}$ & -3.825$^{+0.372}_{-0.544}$ & -3.906$^{+0.372}_{-0.546}$ & -4.025$^{+0.372}_{-0.545}$\\ 
8.625 & -3.141$^{+0.270}_{-0.404}$ & -3.201$^{+0.272}_{-0.451}$ & -3.434$^{+0.273}_{-0.468}$ & -3.557$^{+0.274}_{-0.476}$ & -3.643$^{+0.273}_{-0.481}$ & -3.804$^{+0.273}_{-0.483}$ & -3.872$^{+0.373}_{-0.545}$ & -3.888$^{+0.373}_{-0.544}$ & -3.987$^{+0.372}_{-0.546}$ & -4.126$^{+0.371}_{-0.548}$\\ 
8.875 & -3.418$^{+0.271}_{-0.419}$ & -3.367$^{+0.272}_{-0.459}$ & -3.523$^{+0.273}_{-0.471}$ & -3.684$^{+0.274}_{-0.479}$ & -3.755$^{+0.274}_{-0.481}$ & -3.836$^{+0.276}_{-0.482}$ & -4.007$^{+0.373}_{-0.548}$ & -4.048$^{+0.373}_{-0.547}$ & -4.052$^{+0.373}_{-0.547}$ & -4.166$^{+0.374}_{-0.546}$\\ 
9.125 & -3.711$^{+0.271}_{-0.434}$ & -3.554$^{+0.273}_{-0.466}$ & -3.644$^{+0.273}_{-0.474}$ & -3.780$^{+0.275}_{-0.481}$ & -3.864$^{+0.275}_{-0.483}$ & -3.954$^{+0.274}_{-0.485}$ & -4.024$^{+0.373}_{-0.548}$ & -4.201$^{+0.372}_{-0.550}$ & -4.239$^{+0.374}_{-0.550}$ & -4.273$^{+0.375}_{-0.548}$\\ 
9.375 & -4.087$^{+0.272}_{-0.456}$ & -3.770$^{+0.274}_{-0.479}$ & -3.769$^{+0.272}_{-0.482}$ & -3.868$^{+0.274}_{-0.484}$ & -4.000$^{+0.274}_{-0.487}$ & -4.084$^{+0.276}_{-0.488}$ & -4.155$^{+0.375}_{-0.549}$ & -4.214$^{+0.374}_{-0.550}$ & -4.379$^{+0.374}_{-0.552}$ & -4.483$^{+0.375}_{-0.554}$\\ 
9.625 & -4.659$^{+0.273}_{-0.481}$ & -4.072$^{+0.273}_{-0.487}$ & -3.944$^{+0.275}_{-0.485}$ & -3.999$^{+0.275}_{-0.487}$ & -4.166$^{+0.280}_{-0.490}$ & -4.253$^{+0.272}_{-0.492}$ & -4.337$^{+0.372}_{-0.554}$ & -4.396$^{+0.377}_{-0.553}$ & -4.448$^{+0.373}_{-0.554}$ & -4.639$^{+0.378}_{-0.553}$\\ 
9.875 & -5.158$^{+0.274}_{-0.491}$ & -4.420$^{+0.276}_{-0.493}$ & -4.185$^{+0.274}_{-0.491}$ & -4.199$^{+0.275}_{-0.491}$ & -4.355$^{+0.276}_{-0.495}$ & -4.489$^{+0.279}_{-0.495}$ & -4.541$^{+0.375}_{-0.558}$ & -4.609$^{+0.376}_{-0.559}$ & -4.668$^{+0.375}_{-0.558}$ & -4.758$^{+0.378}_{-0.558}$\\ 
10.125 & -5.792$^{+0.280}_{-0.499}$ & -4.887$^{+0.283}_{-0.500}$ & -4.550$^{+0.277}_{-0.499}$ & -4.547$^{+0.278}_{-0.500}$ & -4.786$^{+0.278}_{-0.503}$ & -4.998$^{+0.281}_{-0.504}$ & -5.188$^{+0.381}_{-0.565}$ & -5.268$^{+0.378}_{-0.566}$ & -5.340$^{+0.379}_{-0.566}$ & -5.434$^{+0.383}_{-0.566}$\\ 
10.375 & -6.488$^{+0.281}_{-0.503}$ & -5.466$^{+0.279}_{-0.504}$ & -5.069$^{+0.282}_{-0.504}$ & -5.096$^{+0.281}_{-0.506}$ & -5.236$^{+0.281}_{-0.506}$ & -5.348$^{+0.280}_{-0.507}$ & -5.491$^{+0.382}_{-0.567}$ & -5.667$^{+0.376}_{-0.567}$ & -5.747$^{+0.381}_{-0.568}$ & -5.874$^{+0.381}_{-0.568}$\\ 
10.625 & -7.142$^{+0.274}_{-0.506}$ & -5.895$^{+0.288}_{-0.505}$ & -5.431$^{+0.280}_{-0.506}$ & -5.488$^{+0.280}_{-0.508}$ & -5.657$^{+0.285}_{-0.508}$ & -5.792$^{+0.278}_{-0.509}$ & -5.895$^{+0.375}_{-0.568}$ & -6.070$^{+0.380}_{-0.568}$ & -6.223$^{+0.380}_{-0.569}$ & -6.334$^{+0.379}_{-0.569}$\\ 
10.875 & -7.809$^{+0.282}_{-0.508}$ & -6.370$^{+0.297}_{-0.506}$ & -5.828$^{+0.287}_{-0.507}$ & -5.906$^{+0.289}_{-0.509}$ & -6.104$^{+0.282}_{-0.509}$ & -6.263$^{+0.284}_{-0.510}$ & -6.401$^{+0.380}_{-0.569}$ & -6.501$^{+0.378}_{-0.569}$ & -6.695$^{+0.383}_{-0.569}$ & -6.878$^{+0.385}_{-0.57}$\\ 
11.125 & -8.540$^{+0.283}_{-0.509}$ & -6.890$^{+0.288}_{-0.507}$ & -6.249$^{+0.282}_{-0.508}$ & -6.337$^{+0.282}_{-0.509}$ & -6.560$^{+0.285}_{-0.510}$ & -6.744$^{+0.281}_{-0.510}$ & -6.902$^{+0.377}_{-0.57}$ & -7.034$^{+0.386}_{-0.57}$ & -7.142$^{+0.379}_{-0.57}$ & -7.359$^{+0.379}_{-0.57}$\\ 
11.375 & -9.335$^{+0.297}_{-0.510}$ & -7.465$^{+0.299}_{-0.508}$ & -6.721$^{+0.282}_{-0.508}$ & -6.800$^{+0.289}_{-0.510}$ & -7.038$^{+0.276}_{-0.510}$ & -7.237$^{+0.279}_{-0.510}$ & -7.405$^{+0.379}_{-0.57}$ & -7.564$^{+0.377}_{-0.57}$ & -7.688$^{+0.377}_{-0.57}$ & -7.841$^{+0.381}_{-0.57}$\\ 
11.625 & -10.285$^{+0.300}_{-0.511}$ & -8.106$^{+0.298}_{-0.509}$ & -7.256$^{+0.284}_{-0.509}$ & -7.315$^{+0.283}_{-0.510}$ & -7.560$^{+0.277}_{-0.510}$ & -7.767$^{+0.279}_{-0.511}$ & -7.944$^{+0.373}_{-0.57}$ & -8.104$^{+0.379}_{-0.57}$ & -8.244$^{+0.379}_{-0.57}$ & -8.403$^{+0.378}_{-0.57}$\\ 
    \end{tabular}}}
    \label{tab:LCII_zall}
\end{table*}

\section{\spr{} IR luminosity function}\label{sec:LIR}

The \spr{} simulation uses as input the redshift evolution of the IR LF of spirals, SB, SF-AGN, SB-AGN, AGN1 and AGN2. In particular, we assumed that the redshift evolution of the characteristic luminosity ($\rm L^{*}$) and the characteristic density ($\Phi^{*}$) of each IR LF, described by a modified Schechter function \citep{Saunders1990}, are described as:
\begin{equation}\label{eq:zevol}
\begin{array}{l}
    \Phi^{*} \propto \begin{cases} (1+z)^{k_{\rho,1}}, & \mbox{if } z\leq z_{\rho} \\ (1+z)^{k_{\rho,2}}, & \mbox{if } z_{\rho}<z<3 \\
    (1+z)^{k_{\Phi}}, & \mbox{if } z \geq 3 \end{cases} \\
    L^{*} \propto \begin{cases} (1+z)^{k_{L,1}}, & \mbox{if } z\leq z_{L} \\ (1+z)^{k_{L,2}}, & \mbox{if } z_{L}<z<3 \\
    \mbox{constant}, & \mbox{if } z_{L} \geq 3 \end{cases}
\end{array}
\end{equation}
For more details on the fit we refers to \citet{Gruppioni2013} and \citetalias{Bisigello2021}, while for the value associated to each parameter we refers to Table 1 in \citetalias{Bisigello2021}.\par
\begin{figure}
    \centering
    \includegraphics[width=\linewidth,keepaspectratio]{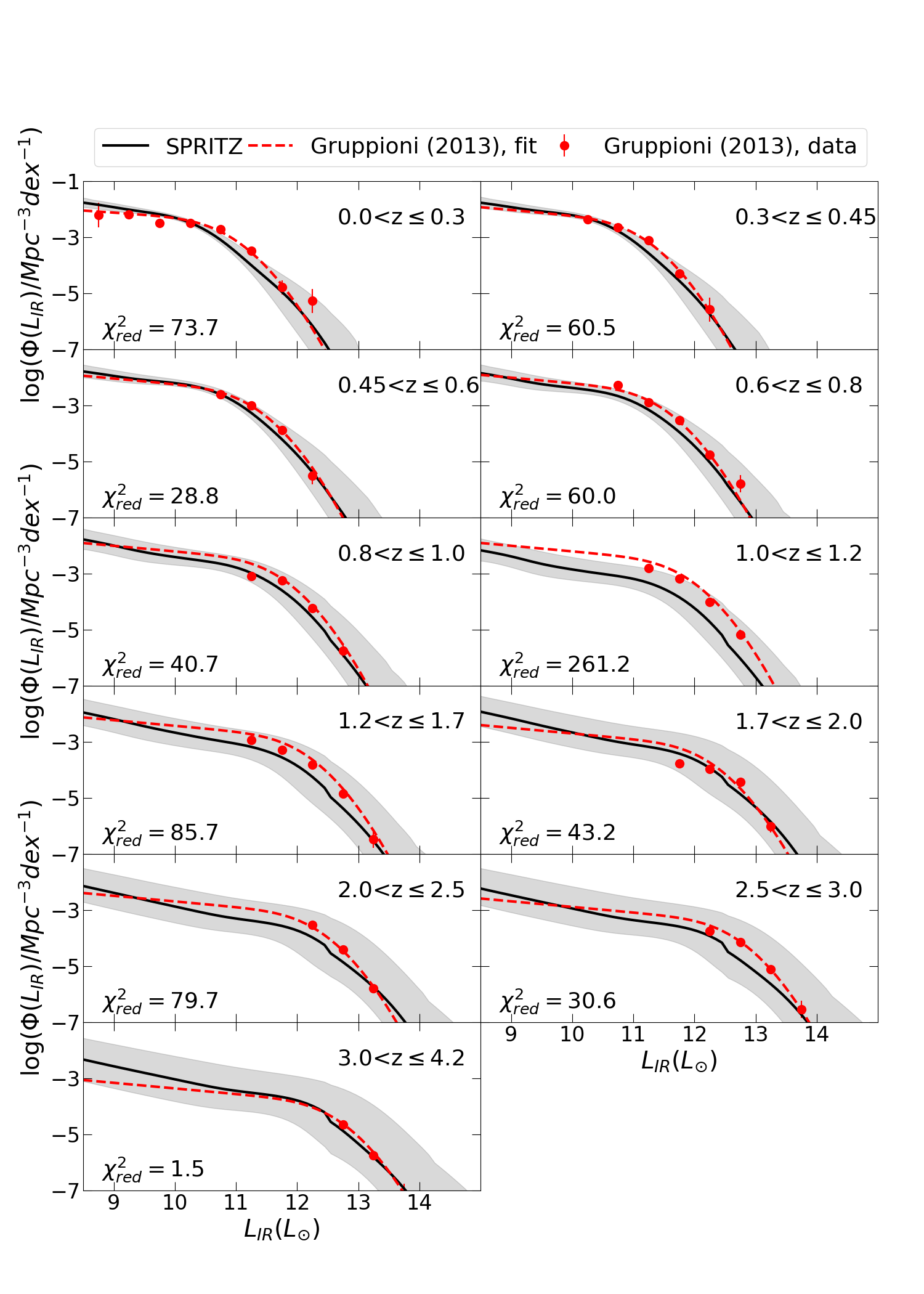}
    \caption{Comparison between the the total IR LF in \spr{} (\textit{black solid lines}) and the total IR LF observed by \hers (\textit{red circles}). We also report the best fits (\textit{red dashed lines}) for each redshift bins, as derived by \citet{Gruppioni2013}. Grey shaded areas indicate the 1$\sigma$ uncertainties associate to the \spr{} IR LF. In each panel we also report the reduced $\chi^{2}$ associated to each redshift bin.}
    \label{fig:LFIR_spritz}
\end{figure}

This parametric description, however, does not describe equally well every redshfit interval, as visible in Figure \ref{fig:LFIR_spritz}. For example, at z=1-1.2 the \spr{} IR LF, even if still consisted with uncertainties, show an underestimation ($\sim$ 0.5-0.8 dex) when compared with the observed one. This may be the cause of some of the discrepancies observed with CO LF (Section \ref{sec:COobs}).

\end{appendix}
\end{document}